\documentclass[10.8pt,twocolumn]{article}
\usepackage{scicite}

\usepackage{times}
\usepackage{amsmath,amssymb}
\usepackage[output-decimal-marker={,}]{siunitx}
\usepackage{braket}
\usepackage{graphicx,caption}
\usepackage{floatrow}
\usepackage{multicol}
\usepackage{wrapfig}
\usepackage{sidecap} 
\usepackage{xcolor}
\newenvironment{sciabstract}{%
\begin{quote} \bf}
{\end{quote}}

\onecolumn
\title{Nontopological zero-bias peaks in full-shell nanowires induced by flux tunable Andreev states}

\author
{Marco Valentini,$^{1\ast}$ Fernando Pe\~naranda,$^2$ Andrea Hofmann,$^1\dagger$ Matthias Brauns,$^1\ddagger$ \\ Robert Hauschild,$^1$ Peter Krogstrup,$^3$ Pablo San-Jose,$^2$ Elsa Prada,$^{2,4}$ \\ Ram{\'o}n Aguado,$^{2\ast}$ Georgios Katsaros,$^{1\ast}$\\
\\
\normalsize{$^{1}$Institute of Science and Technology Austria}\\
\normalsize{Am Campus 1, 3400 Klosterneuburg, Austria}\\
\normalsize{$^{2}$Instituto de Ciencia de Materiales de Madrid (ICMM),} \\ \normalsize{Consejo Superior de Investigaciones Cient\'{i}ficas (CSIC)}\\
\normalsize{Sor Juana In\'{e}s de la Cruz 3, 28049 Madrid, Spain.}\\
\normalsize{$^{3}$Microsoft Quantum Materials Lab and Center for Quantum Devices, Niels Bohr Institute}\\
\normalsize{University of Copenhagen, Kanalvej 7, 2800 Kongens Lyngby, Denmark}\\
\normalsize{$^{4}$Departamento de F\'isica de la Materia Condensada, Condensed Matter Physics Center (IFIMAC)} \\ \normalsize{and Instituto Nicol\'as Cabrera, Universidad Aut\'onoma de Madrid}\\
\normalsize{E-28049 Madrid, Spain}\\
\\
\normalsize{$^\ast$Corresponding author. E-mail:} \\ \normalsize{marco.valentini@ist.ac.at; raguado@icmm.csic.es; georgios.katsaros@ist.ac.at.}
\\
\normalsize{$^\dagger$ Present address: Swiss Re Insurance Company Ltd, Mythenquai 50/60, 8002 Zürich, Switzerland.}
\\
\normalsize{$^\ddagger$ Present address: XARION Laser Acoustics GmbH, Ghegastrasse 3, 1030 Vienna, Austria.}
}

\date{}

\begin{document} 



\maketitle


\begin{sciabstract}
A semiconducting nanowire fully wrapped by a superconducting shell has been proposed as a platform for obtaining Majorana modes at small magnetic fields. In this study, we demonstrate that the appearance of subgap states in such structures  is actually governed by the junction region in tunneling spectroscopy measurements, and not the full-shell nanowire itself. Short tunneling regions never show subgap states, whereas  longer junctions always do. This can be understood in terms of quantum dots forming in the junction and hosting Andreev levels in the Yu-Shiba-Rusinov regime. The intricate magnetic field dependence of the Andreev levels, through both the Zeeman and Little-Parks effects, may result in robust zero-bias peaks, features that could be easily misinterpreted as originating from Majorana zero modes, but are unrelated to topological superconductivity.
\end{sciabstract}

\twocolumn

\twocolumn[
{\includegraphics[width=\textwidth]{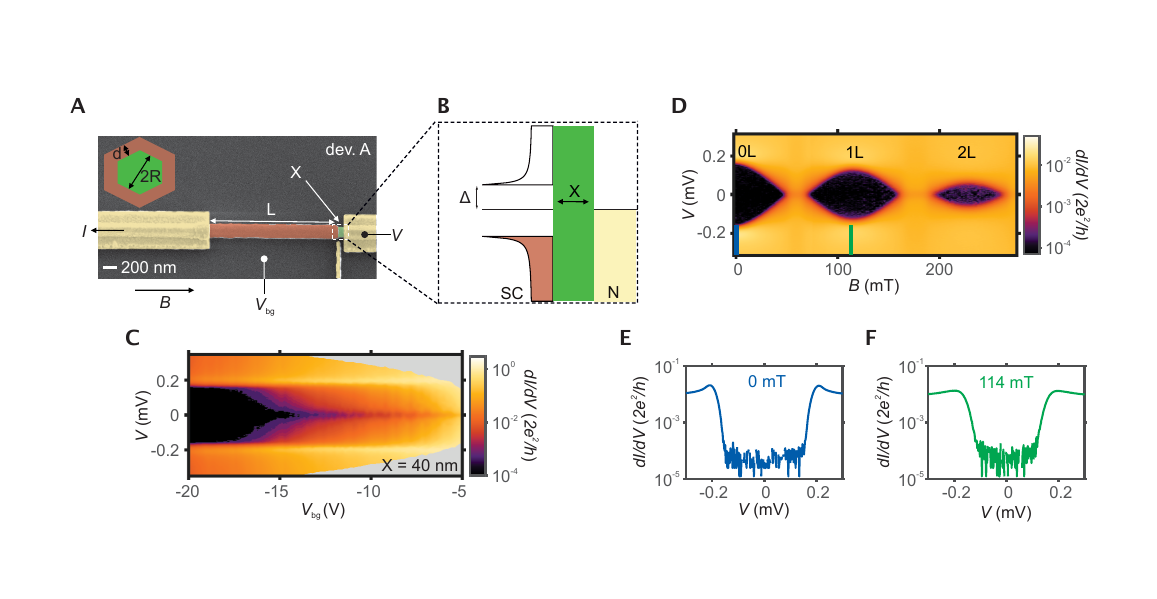}
\captionof{figure}{\textbf{Tunneling spectroscopy of a short-junction device.} \small{ (\textbf{A})  False-color scanning electron micrograph of device A. A tunnel junction of length $X\approx 40\,\si{nm}$ created in the bare InAs NW (green) is formed between the normal contact (yellow) and the proximitized full-shell NW (red), of length $L$. It can be tuned by the overall backgate $V_{\textrm{bg}}$ (sidegate is grounded). The electrochemical potential $\mu$ inside the full-shell cannot be gated because of the Al screening, hence $V_{\textrm{bg}}$ affects only the tunnel region. 
A magnetic field $B$ is applied parallel to the NW. The inset shows the hexagonal NW cross section with a semiconducting core radius $R\approx 55 \, \si{nm}$ and a shell thickness $d\approx 30 \, \si{nm}$. (\textbf{B})  Sketch of a superconductor(SC)-tunnel barrier (TB)-normal metal (N) junction.  Our measurements show that no QD is formed in the bare InAs NW for $X\lesssim 100\,\si{nm}$. (\textbf{C})  Differential conductance $dI/dV$ plotted on a logarithmic scale as a function of source-drain bias voltage $V$ and $V_{\textrm{bg}}$, which tunes the tunnel barrier transparency.  (\textbf{D}) $dI/dV$ as a function of $V$ and $B$ for device A at $V_{\textrm{bg}} = -17 \, \si{V}$. LP oscillations are observed, in which superconducting lobes (denoted 0L, 1L and 2L) are separated by regions where superconductivity is completely suppressed. From the shape of the lobes, one can extract the NW dimensions $R\approx 64$ nm and $d \approx 24$ nm, in good agreement with the nominal values mentioned above) and the coherence length ($\xi \approx 160$ nm), as shown in Fig. S5. (\textbf{E} and \textbf{F}) Line-cuts taken from (\textbf{D}) at the center of 0L and 1L, respectively. In both cases the gap is hard.}}\label{fig:figure1}\par\bigskip}]

\begin{figure*}[h]
\includegraphics[scale=1]{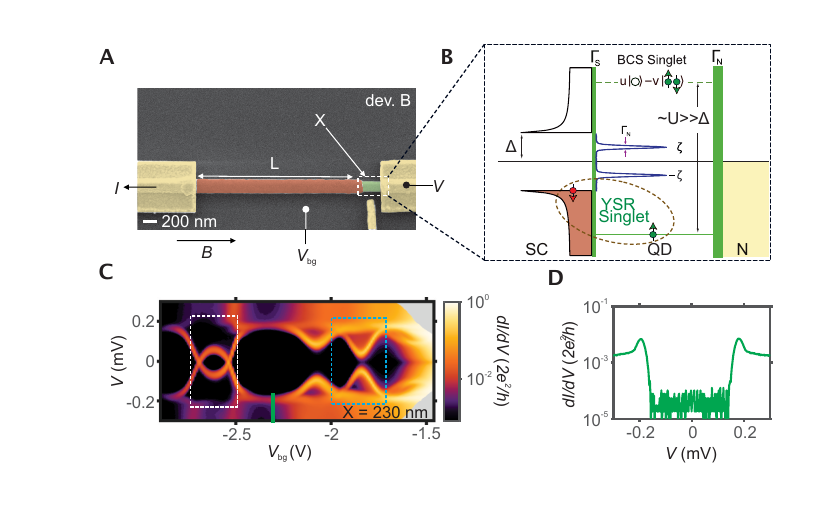}%
\caption{\label{fig:figure2} \textbf{Tunneling spectroscopy of a long-junction device.} \small{(\textbf{A})  False-color scanning electron micrograph of device B.  The tunnel region of length $X\approx 240 \, \si{nm}$ accommodates a QD with charging energy $U \approx  2.5 \, \si{meV}$, as measured from the stability diagram taken at a magnetic field that suppresses superconductivity. (\textbf{B})  Sketch of a superconductor (SC)-qunatum dot (QD)-normal metal (N) junction, showing all of the physical energy scales involved in this problem, namely, $U$, the superconducting gap $\Delta$, and the tunnel rates $\Gamma_S/\hbar$ and $\Gamma_N/\hbar$ into the SC and the metal ($\Gamma_{S,N}$ have energy units). For the devices indicated here $U\gg\Delta$ and the first excited state over the doublet GS at odd occupation is a Kondo-like singlet between the unpaired spin in the QD and the quasiparticles in the SC [YSR singlet]. This excitation creates subgap states inside $\Delta$ at energies $\pm\zeta$. Another possible excitation is the BCS-like singlet superposition of even charge states in the QD, which, however, is much higher in energy (order $\sim U$). (\textbf{C}) $dI/dV$ plotted on a logarithmic scale as a function of  $V$ and $V_{\textrm{bg}}$. White and blue dashed boxes indicate a doublet
and a singlet GS region, respectively. (\textbf{D}) Line cut extracted from (C) at  $V_{\textrm{bg}}=-2.3 \, \si{V}$. Although, this device supports ABSs, a hard gap can be observed for certain gate voltage values.}}
\end{figure*}

The superconducting Bardeen–Cooper–Schrieffer (BCS) density of states (DOS) is characterized by a gap $\Delta$ around the Fermi energy that prevents quasiparticle excitations at energies below it. However, in systems with spatially-inhomogeneous pairing potentials, such as weak links between two superconductors (SCs), the DOS contains states inside the gap known generically as Andreev bound states (ABSs) \cite{Sauls:PTRSA18}. Such ABSs have been observed in e.g. carbon nanotubes ~\cite{pillet2010andreev,Eichler:PRL07}, superconducting atomic point contacts \cite{Bretheau:N13}, graphene \cite{Dirks:NP11} and hybrid semiconducting-superconducting systems based on nanowires (NWs) \cite{Lee:PRL12,Chang:PRL13,Lee:NN14,PhysRevB.94.064520,Lee:PRB17,Grove-Rasmussen:NC18,Su:PRL18,Junger:CP19,junger2020,PhysRevB.101.235315}.
ABSs are notable in their own right, as Andreev qubits \cite{Janvier:S15,Hays-FatemiNP2020}, and for the rich physics they offer; however, the intense research activity of the past few years in the hybrid NW platform has arguably been motivated by the prediction that a topological SC state with Majorana zero modes (MZMs) can be engineered out of them \cite{lutchyn_majorana_2010,oreg_helical_2010} [see also \cite{Aguado:RNC17,Lutchyn:NRM18,Prada:NRP20} for recent reviews]. 
Despite the fact that several experiments in such platforms have reported on signatures compatible with MZMs \cite{mourik_signatures_2012,albrecht_exponential_2016,Deng:S16,Nichele:PRL17,Zhang:NC17,Gul:NN18}, the Majorana interpretation has been challenged because zero-energy ABSs, in the absence of an underlying topological state, can mimic MZMs \cite{Prada2012,Kells2012,liu2017andreev,Moore2018,PhysRevB.98.245407,Vuik2019,Avila2019,Chen2019,pan2020,cayao2015sns,Prada:NRP20}.

\begin{figure*}[h]
\includegraphics[scale=1]{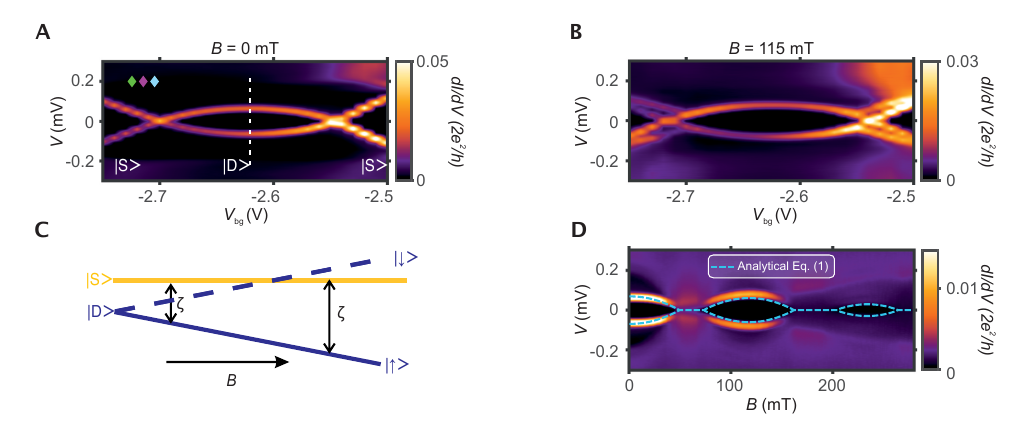}%
\caption{\label{fig:figure3} \textbf{ABSs in the LP regime with a doublet GS.} \small{ (\textbf{A}) Plot showing a zoomed-in view of the region denoted by the white box in  (Fig. 2C). for device B. From $V_{\textrm{bg}} = -2.70 \, \si{V}$ to $V_{\textrm{bg}} = -2.55 \, \si{V}$ the QD is in a doublet GS. Colored diamonds correspond to images in (Fig. 4E). (\textbf{B})  Same as (A) but for $B = 115 \, \si{mT}$, at the center of the first lobe, see (D). (\textbf{C}) Schematics showing the evolution of the ABSs with a doublet GS when a magnetic field $B$ is switched on. (\textbf{D}) $dI/dV$ as a function of  $V$ and $B$ for $V_{\textrm{bg}} = -2.62 \, \si{V}$, at the center of the doublet GS [white dashed line in (A)]. The blue dashed curve corresponds to the analytical level positions given in Eq. \ref{YSR-Zeeman}, showing that these ABSs are of the YSR type [where we use $U=2.5$ meV (see (Fig. 2) caption), and $\Gamma_S=0.8$ meV$ =0.34\,U$ is inferred from the position of the ABSs at $B=0$]. Line-cuts taken at different magnetic fields as well as the Kondo effect observed in the destructive regimes are discussed in \cite{supplement}.}}
\end{figure*}

\begin{figure*}[h]
\includegraphics[scale=0.97]{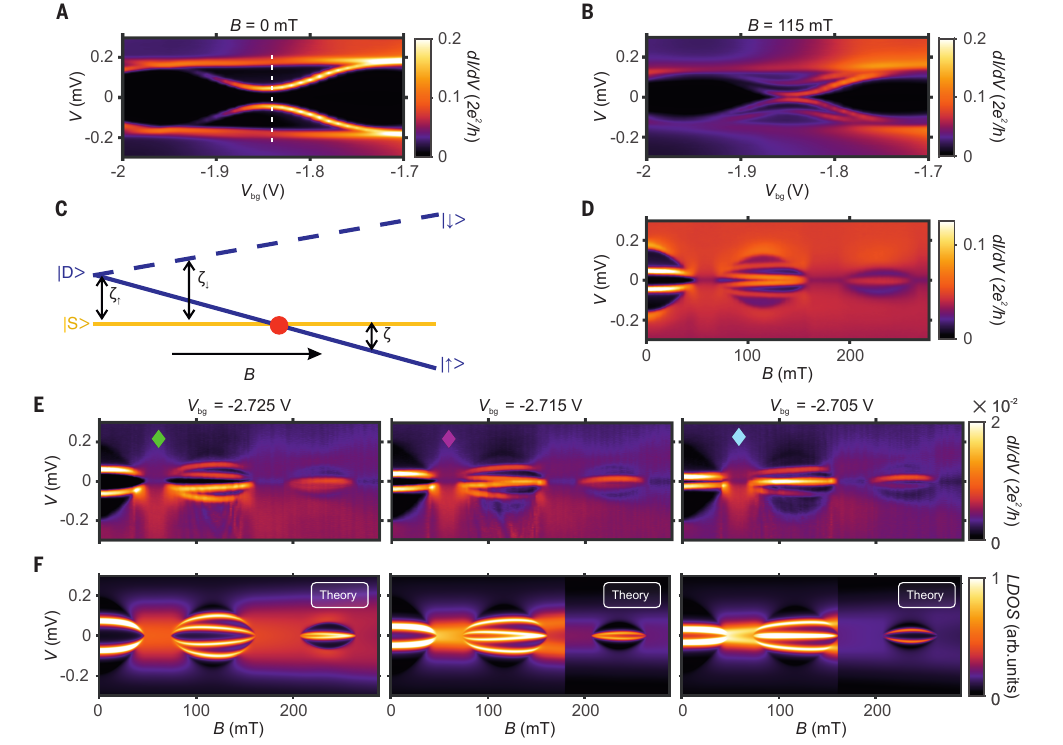}%
\caption{\label{fig:figure4} \textbf{ABSs in the LP regime with a singlet GS.} 
\small{(\textbf{A}) Plot showing a zoomed-in view of the region denoted by the blue dashed box in Fig. 2C around a singlet GS for device B. (\textbf{B}) Same as (A) but at finite magnetic field $B = 115 \, \si{mT}$. The ABSs split and the splitting depends on $V_{\textrm{bg}}$. 
(\textbf{C}) Schematics of the ABS behavior with a singlet GS when a magnetic field $B$ is switched on.
(\textbf{D}) $dI/dV$ as a function of  $V$ and $B$ for $V_{\textrm{bg}} = -1.84 \, \si{V}$ [dashed white line in (A)]. A ZBP starts at the end of the 1L and persists throughout all of the 2L,  thus extending thus for more than 100 mT. 
(\textbf{E}) $dI/dV$ plotted on a logarithmic scale as a function of  $V$ and $B$ for  $V_{\textrm{bg}} = -2.725$, $-2.715$ and $-2.705\, \si{V}$ (left to right); see colored diamonds in (Fig. 3A). By changing the ABS position at zero field, it is possible to create a situation in which the ABSs form a ZBP throughout the 1L. (\textbf{F}) Numerical simulation
of the LDOS versus $V$ and $B$ in a QD-SC system, with the SC in the destructive LP regime and the QD in a singlet GS, for three different gate configurations (parameters in Table S1). The QD-SC system is modeled as a superconducting Anderson model, with finite level broadening induced by $\Gamma_N$ to qualitatively match the experimental $dI/dV$ peak widths.}}
\end{figure*}

Recently, it has been argued that full-shell NWs \cite{chang_hard_2015} threaded by a magnetic flux $\phi=AB$, with $A$ the cross section of the NW and $B$ an external axial magnetic field, are an alternative platform for realizing topological superconductivity \cite{Vaitiekenaseaav3392}. The full-shell geometry has the great advantage that a topological phase can be induced at known and relatively low magnetic fields; typically when $\phi$ is close to one flux quantum ${\phi}_0 = h/2e$ [the precise range depends on the geometry of the full-shell NW \cite{fern2019evenodd}; here $h$ is the Planck's constant and $e$ is the electron's charge]. As shown by experiments \cite{vaitieknas2019anomalous, Vaitiekenaseaav3392, sabonis2020destructive}, InAs/Al full-shell NWs exhibit flux tunability of the superconducting gap, including its complete destruction and reemergence, owing to the Little-Parks (LP) effect \cite{schwiete_2010,shah_2007}. The LP effect is accompanied by zero bias peaks (ZBPs) in tunneling conductance, appearing in reentrant superconducting regions around $\phi\sim{\phi}_0$ and have been interpreted as MZMs \cite{Vaitiekenaseaav3392}. In this study, we use similar InAs/Al full-shell NWs and perform tunneling spectroscopy measurements to investigate the role of the tunneling junction (green region of length $X$ in Fig. 1A) on the subgap spectra. Details on the measured NWs can be found in the supplementary ''materials and methods'' \cite{supplement}. We use three experimental knobs, the magnetic flux, the junction length $X$ and the global backgate voltage. Data from more than 40 devices demonstrate that for $X\lesssim 100\,\si{nm}$, the bare InAs NW acts as a tunnel barrier, whereas for $X \gtrsim 150 \, \si{nm}$, a quantum dot (QD) is formed in the tunnel junction region. Data for $100\,\si{nm}\lesssim X\lesssim 150\,\si{nm}$ show that in this range a QD may or may not form; we therefore do not work in this tunnel-junction length regime. 
The main text discusses data from five devices, which we refer to as A, B, C, D and E. Data from eight more devices are presented in the supplementary material \cite{supplement}. All measurements have been performed in a dilution refrigerator with a base temperature of $20\,\si{mK}$.

\section*{Short-junction devices}
We initially focus on the short-junction devices for which the bare InAs NW plays the role of a tunnel barrier (Fig. 1B).
Figure 1C displays the measured differential conductance at zero magnetic field of device A with a short tunnel junction of $X\approx 40\,\si{nm}$. With increasingly negative $V_{\textrm{bg}}$, the tunneling conductance at source-drain voltages $V$ below the superconducting gap $\Delta$ decreases and reveals a hard gap of size $\Delta\approx200 \, \si{\micro eV}$. The line trace shown in Fig. 1E confirms the absence of subgap states and a hard gap with a subgap conductance suppressed by a factor of  $\sim 300$ relative to the above-gap conductance. Similar tunneling spectroscopy 
data have been observed for all devices with a junction length X $ < 100 \, \si{nm}$ (Fig. S6). The hardness of the gap and the absence of subgap features make the short junction devices the best candidates for an unambiguous detection of MZM signatures.

We now discuss transport spectroscopy data as a function of an external parallel magnetic field $B$. Figure 1D shows the $dI/dV$ evolution ($I$, current; $V$, voltage) as a function of $B$ at a fixed value of $V_{\textrm{bg}}$ for device A. The  modulation of $\Delta$ with $B$ is the result of the destructive LP effect \cite{LP_1962,vaitieknas2019anomalous,liu2001destruction}, with regions where the gap is completely suppressed and subsequently regenerated (hereafter we label the regions with finite gaps as zeroth lobe, first lobe, and so on). Theory predicts the observation of a ZBP in the first lobe owing to the formation of MZMs \cite{Vaitiekenaseaav3392,fern2019evenodd}.
Notwithstanding these predictions, none of the nine short junction devices reported here (Fig. 1 and Fig. S6) exhibit a ZBP (or any other subgap state) as the magnetic field increases. Instead, a hard gap slightly smaller than at zero field is observed in the first lobe (Fig. 1F).
Possible explanations for the absence of ZBPs in the first lobe include the NWs always being topologically trivial owing to the renormalization of the semiconductor properties caused by strong coupling to the SC \cite{PhysRevB.97.165425,PhysRevX.8.031040}, or lack of proper tuning into a topological phase \cite{fern2019evenodd,PhysRevB.99.161118,PhysRevB.101.054515}.
The described short-junction results seemingly contradict a recent experimental report on MZMs in similar NWs \cite{Vaitiekenaseaav3392}. However, the lithographic length of the shown tunnel junction in \cite{Vaitiekenaseaav3392} seems to be larger than $100 \, \si{nm}$ and falls into what we call the long junction regime. In the next section, we discuss how, in this regime, we observe ZBPs of non topological origin owing to QD physics.

\section*{Long-junction devices}

Figure 2A shows a micrograph of device B with $X\approx 240 \, \si{nm}$, for which a QD is formed in the tunnel junction as sketched in Fig. 2B. In this QD-SC system, two ground states (GSs) are accessible: a spin doublet, $ \ket{D}$, with spin $\frac{1}{2}$, and a spin singlet, $ \ket{S}$, with spin 0.
Whether the GS is a doublet or a singlet is determined by the interplay between the Andreev processes at the SC-QD interface and Coulomb blockade (CB), with charging energy $U$ in the QD. CB enforces a one by one electron filling, making it possible to have odd electron occupations with a doublet GS~\cite{PhysRevB.68.035105,PhysRevB.79.224521,tarucha2010,ramon2015}. The coupling to the SC, $\Gamma_S$, on the other hand, favors  a singlet GS. Its physical nature depends on the ratio $\Delta/U$. In the large  $\Delta/U$ limit, the coupling to the SC mainly induces local superconducting correlations in the QD that lead to Bogoliubov-type singlets, 
which are superpositions of the empty $\ket{0}$ and doubly-occupied $\ket{\uparrow\downarrow}$ states in the QD. In the opposite, small $\Delta/U$ limit, the unpaired spin in the QD couples to the quasiparticles in the SC, see Fig. 2B, with an exchange interaction $J\sim 2\Gamma_S/U$. This exchange interaction creates so-called Yu-Shiba-Rusinov (YSR) singlets, the superconducting counterpart of Kondo singlets  \cite{Yu:APS65,Shiba:PTP68,Rusinov:SPJ69,RevModPhys.78.373}. For small $J \ll 1$, the GS is a doublet and the YSR singlets occur as discrete ABS excitations near the edge of the SC gap. A larger $J$, however, moves these excitations to subgap energies and, eventually, may induce zero-energy crossings when $J \approx 1$, signaling a quantum phase transition (QPT) in which the YSR singlet becomes the new GS. Because our experiments are always in the large $U$ limit (even for the largest gap at $\phi=0$, we always have $U\gg\Delta$), the YSR regime \cite{Chang:PRL13,Lee:NN14,PhysRevB.94.064520,Lee:PRB17,PhysRevB.82.245108,ramon2015,PhysRevB.92.235422} is the relevant one [for a full theoretical discussion of all physical regimes, see \cite{supplement}]. Transitions between the GS and the first excited state of the system, i.e., between a doublet and a singlet state or vice versa, manifest in transport spectroscopy as a subgap resonance at voltage $V=\zeta$ and its electron-hole-symmetric partner at $V=-\zeta$ ($\zeta$, energy difference between excited state and GS). Furthermore, changes in the parity of the GS of the system appear as points in parameter space (here $V_{\textrm{bg}}$ or $B$) where $\zeta$ changes sign (signalled by the crossing of ABSs at zero energy) ~\cite{Chang:PRL13,Lee:NN14,PhysRevB.94.064520,Lee:PRB17}.

Figure 2C displays the measured differential conductance of the long-junction device B at zero magnetic field. As explained above, a symmetric pair of conductance peaks appears at $|V|<\Delta$. $V_{\textrm{bg}}$ can tune the position of these subgap states that move up and down in energy and can even induce zero-energy crossings that form a characteristic eye-shaped loop (white dashed rectangle), which is the superconducting analog of an odd-occupation CB valley. In the middle of this valley, the system behaves as a spin-$1/2$ impurity, which induces YSR physics.
Far from the odd valleys, $V_{\textrm{bg}}$ tunes the ABSs to higher energies (green vertical line) where they merge and disappear into the continuum of states. In this region, the line trace in Fig. 2D again reveals a hard gap, with a subgap conductance suppressed by a factor of  $\sim 90$ relative to the above-gap conductance.

\section*{Magnetic field evolution of ABSs in the YSR limit}

Next, we consider a doublet GS, occurring in a $V_{\textrm{bg}}$ region between $-2.70$ and $-2.55\, \si{V}$ (white box in Fig. 2C). Figure 3A displays a zoomed-in view of this region, illustrating that the gap is clean apart from a single pair of ABSs at energies $\pm\zeta$ . 
At finite magnetic fields, $|\zeta|$ slightly increases in the doublet region compared to zero field compared with its value at zero field (Fig. 3B). The increasing energy results from the decreasing doublet GS energy owing to the Zeeman effect. The spin-polarized doublet states change their energy with $B$ by the Zeeman energy $\pm V_Z=\pm g\mu_BB/2$ (with $g$ and $\mu_B$ being the $g$-factor and the Bohr's magneton, respectively), whereas the excited singlet energy remains unaffected (Fig. 3C). Notably, because the GS is spin-polarized, there is only one allowed excitation (vertical arrows in Fig. 3C). This excitation results in a pair of ABSs inside the LP lobes (as opposed to the short junction discussed in Fig. 1). This is illustrated in Fig. 3D, where we show the full magnetic field evolution for fixed $V_{\textrm{bg}} = -2.62\, \si{V}$ (dashed line in Fig. 3A). This $B$-field evolution strongly deviates from the linear increase expected for a standard Zeeman effect and stays nearly constant within the zeroth lobe. Because we fix $V_{\textrm{bg}}$ in the middle of the loop (middle of the CB valley), charge fluctuations are substantially suppressed in this configuration and the system should essentially behave as a spin 1/2 coupled to a SC. Indeed, by modelling the system as a CB QD coupled to a SC lead (the so-called superconducting Anderson model), we can write an analytic expression for the ABSs in this large-$U$ limit of the form \cite{supplement}: 
\begin{equation}
\label{YSR-Zeeman}
\zeta=\pm\Delta(\phi)\frac{1-\left(\frac{2\Gamma_S}{U+V_Z}\right)^2}{1+\left(\frac{2\Gamma_S}{U+V_Z}\right)^2},
\end{equation}
where the QD level position is fixed to $\epsilon_0=-U/2$ to describe the spinful odd CB valley.
This equation is  the expression for YSR bound states \cite{RevModPhys.78.373} written in the language of our system and including the external magnetic flux [through both the LP modulation of the superconducting gap $\Delta(\phi)$  and the Zeeman effect $V_Z$ in the QD]. The dashed blue curves in Fig. 3D are calculated with this analytical expression [where the charging energy $U$ is extracted from the experimental stability diagram in the normal state, $\Delta(\phi)$ is fitted to the experimental LP gap evolution, and the remaining free parameter $\Gamma_S$ is fixed by the experimental position of the ABS at zero magnetic field].
The good agreement between Eq. \eqref{YSR-Zeeman} and the experiment demonstrates that, indeed, our ABSs are YSR singlets [for comparison, the excitations to the BCS-like singlet $\ket{S}=\ket{0}- \ket{\uparrow\downarrow}$ would occur at a much higher energy, of order $\zeta\approx \pm (U-\Gamma_S)/2=0.85$ meV$\gg\Delta$].

\section*{Magnetic field evolution of ABSs with a singlet GS: Deep in-gap limit}
We can tune the device to a singlet GS region by making the backgate voltage less negative (see Fig. 2C around $V_{\textrm{bg}} = -1.84 \, \si{V}$; blue dashed rectangle). The excitations from this singlet GS are doublet-like and hence split under a Zeeman field~\cite{Lee:NN14}, as opposed to the previous case. Figure 4A presents a detailed zoomed-in view of device B in this gate region. Figure 4B shows the same scan but for $B = 115\, \si{mT}$ (corresponding to the center of the first LP lobe). Because two spin-polarized excitations from the singlet GS are possible~\cite{Lee:NN14}, two pairs of ABSs are observed: one near the gap center and one closer to the gap edge (denoted as $\zeta_\uparrow$  and  $\zeta_\downarrow$ in Fig. 4C). 
The $B$-field evolution at $V_{\textrm{bg}} = -1.84 \, \si{V}$ (dashed line in Fig. 4A) is displayed in Fig. 4D. In the zeroth lobe, the pair of ABSs neither splits nor moves. By contrast, the ABSs show a clear Zeeman splitting in the first lobe. For increasing magnetic fields, the lowest ABS excitation moves towards zero energy at the end of the first lobe, forming a ZBP. This zero-energy crossing signals a QPT to a spin polarized doublet GS. The ZBP persists throughout the second lobe, as revealed in Fig. 4D. Our theoretical analysis of this regime  
supports this interpretation and explains the absence of a clear Zeeman splitting in  the zeroth lobe (Fig. S3).

We now examine $B$ field driven ZBPs in a different parameter regime: we consider the evolution of in-gap ABSs far from the gap edge at $B=0$ but close to a singlet-doublet transition enabled by varying $V_{\textrm{bg}}$ (see the three colored diamonds in Fig. 3A). Our results are presented in Fig. 4E. The three subpanels show the evolution in magnetic field of the same pair of ABSs but at  slightly different backgate values (Fig. S11). 
As the backgate moves closer to the singlet-doublet zero-energy crossing, the ABS energy decreases. For increasing magnetic fields, this $B=0$ energy influences the particular $B$ value at which the excitation reaches zero energy (compare the panels in Fig. 4E from left to right). 
This gate dependence makes it possible to tune the value of $B$ at which the ZBP emerges. For $V_{\textrm{bg}}=-2.705  \si{V}$ (blue diamond), 
the ZBP appears at the beginning of the first lobe and persists throughout its full extent. The three panels of Fig. 4F show numerical simulations of the local density of states (LDOS) in this regime, whose overall agreement with the tunneling $dI/dV$ spectroscopy is excellent [for technical details on the relation between the two quantities, see \cite{supplement}]. This showcases how ABSs may produce a strong ZBP that persists across a large range of magnetic fields~\cite{Lee:NN14} as a result of their finite line-width and the repulsion from the field-modulated gap edge.
%After the ZBP, the GS changes from singlet to doublet. 
After the formation of the ZBP, the GS changes from singlet to doublet when the field is increased. This spin-polarized YSR doublet again follows Eq. \ref{YSR-Zeeman} \cite{supplement}.

\onecolumn
\begin{figure}[h]
\includegraphics[scale=1]{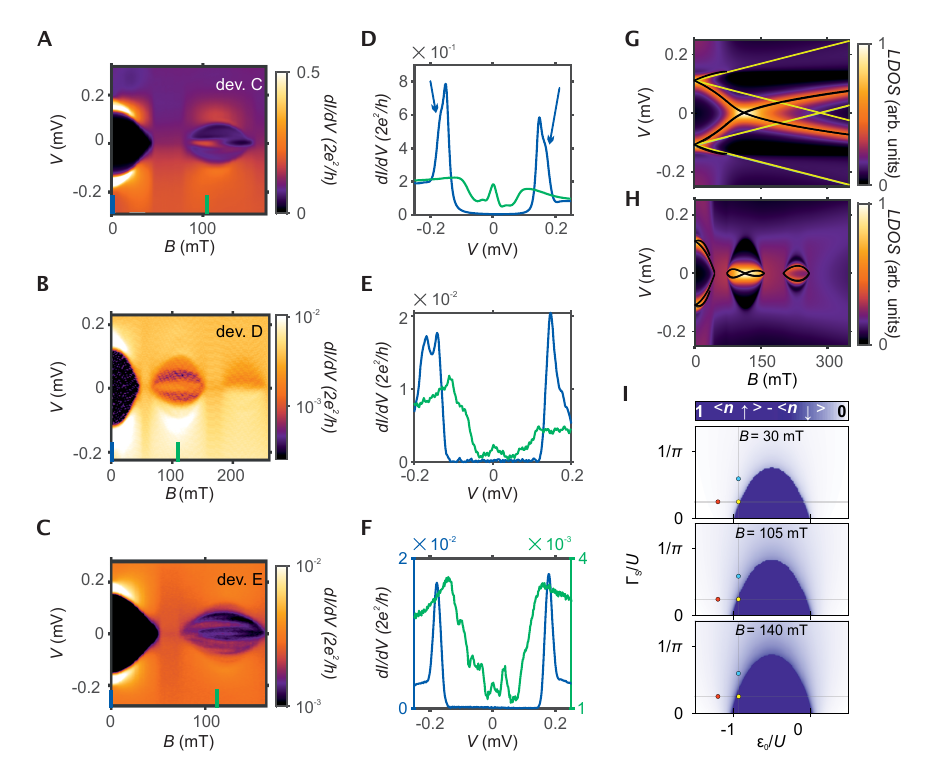}%
\caption{\label{fig:figure5} \textbf{ZBPs arising from ABSs ''hidden'' in the superconducting gap edge.} \small{(\textbf{A} to \textbf{C}) $dI/dV$ for three different devices as a function of  $V$ and $B$ for $V_{\textrm{bg}} = -1.84\, \si{V}$, $-1.27$ and $-1.89\, \si{V}$, respectively. For (B) and (C) a logarithmic scale have been used. ABSs close to the gap edge at zero field, where the GS is in the singlet state, converge to a ZBP in the 1L. (\textbf{D} to \textbf{F}) $dI/dV$ versus $V$ traces extracted from (A-C), at $B = 0\, \si{mT}$ (blue) and at the middle of the 1L at $B = 102$, $112$ and $123\, \si{mT}$ (green), respectively, showing a ZBP. (\textbf{G} and \textbf{H}) Numerical simulation of the LDOS as a function of  $V$ and $B$ for parameters corresponding to device D. In (G) there is a constant pairing equal to $\Delta$, whereas a destructive LP modulation of the BCS gap is considered in (H).
The evolution with $B$ of the Green’s function poles of a superconducting Anderson model are shown in black. For comparison, we also show in (G) the evolution of the ABS with increasing $B$ assuming that only Zeeman splitting is relevant (yellow lines). The observation of the non linear $B$ evolution of the ABSs strongly depends on the individual parameters of the device, namely $\Gamma_{S}$, $U$ and their position at $B=0$.
(\textbf{I}) Phase diagram of the dot-level spin-polarized occupations $\langle n_\uparrow \rangle- \langle n_\downarrow \rangle$ versus $\Gamma_S/U$ and $\epsilon_0/U$ for three different magnetic fields: $B = 30 \, \si{mT}$, $B = 105 \, \si{mT}$, and $B = 140 \, \si{mT}$ [for the model parameters see \cite{supplement}]. $\epsilon_0$ is the dot level energy at zero field. The yellow circle corresponds to the specific configuration in the phase diagram that leads to (H). The singlet-doublet transition line (between white and blue regions) as a function of $B$ exhibits a strong dependence on $\Gamma_S/U$. As the magnetic field increases, a singlet phase (top) becomes a doublet phase (bottom), across a parity crossing located at $B = 105 \, \si{mT}$ (center). The red and blue circles denote similar experimental configurations, but with slightly different gating and coupling to the SC, that never cross the singlet-doublet transition line for the magnetic fields shown.}}
\end{figure}
\twocolumn

\begin{figure*}[h]
\includegraphics[scale=0.97]{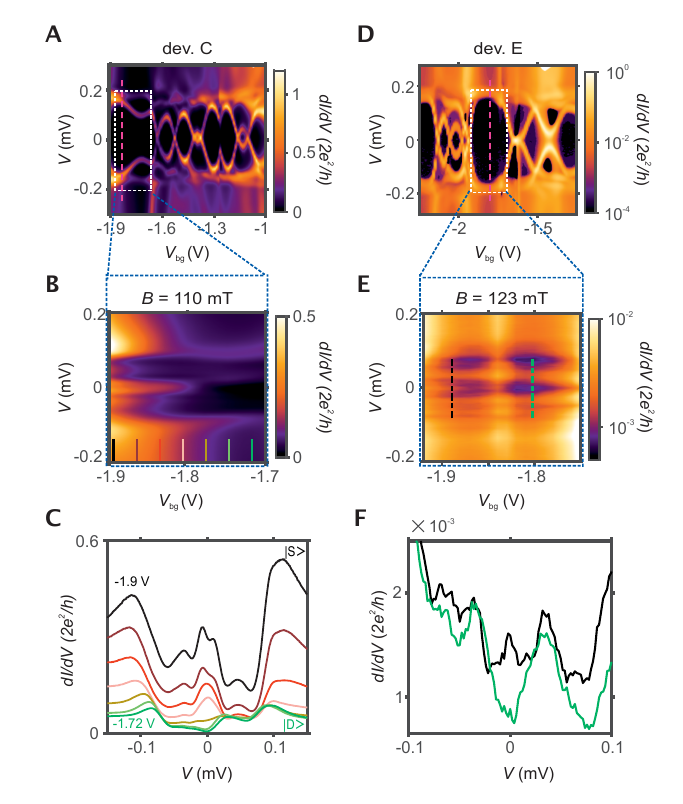}%
\caption{\label{fig:figure6} \textbf{Gate-voltage dependence of ZBPs.} \small{(\textbf{A} and \textbf{D}) $dI/dV$ as a function of $V$ and $V_{\textrm{bg}}$ in the absence of magnetic field for device C and E, respectively. The dashed pink lines indicate where the magnetic field scans shown in Fig. 5, A and C, were taken. (\textbf{B} and \textbf{E}) Zoomed-in views of (A) and (D) but taken at the center of the 1L, i.e. at $B = 110$ and $123 \, \si{mT}$, respectively. For (D) and (E) a logarithmic scale has been used.
(\textbf{C}) Line traces extracted from (B), ranging from $V_{\textrm{bg}} = -1.9\, \si{V}$ (black)  to $V_{\textrm{bg}} = -1.72\, \si{V}$ (dark green) and shifted by $0.25 \frac{e^{2}}{h}$ with respect to each other. At $V_{\textrm{bg}} = -1.8\, \si{V}$ the ABSs cross zero bias and the GS becomes doublet. (\textbf{F}) Line-cuts extracted from (E) at $V_{\textrm{bg}} = -1.89 \, \si{V}$ (black) and  $V_{\textrm{bg}} = -1.8\, \si{V}$ (dark green). A ZBP is present at $V_{\textrm{bg}} = -1.89 \, \si{V}$ but disappears for higher gate voltages.}}
\end{figure*}

\section*{Magnetic field evolution of ABSs with a singlet GS: Near gap edge limit}

Finally, we turn our attention to devices that show ZBPs without an apparent subgap structure in the zeroth lobe.
In Fig. 5 we present data from devices C, D and E, with tunnel junction lengths of $X\approx 290$, $200$ and $300\, \si{nm}$, respectively. 
Figure 5, A to C, depicts the magnetic field dependence of the respective tunneling spectra, with a ZBP emerging in the first lobe after the first LP closing.
Notably, and as opposed to our previous results, 
no obvious signatures of ABSs can be inferred from the spectra of the zeroth lobe. Taken together, all of these spectroscopic features could be interpreted as originating from MZMs. However, as we argue now, these spectra can also be understood in terms of the magnetic field evolution of ABSs.
We note that the $dI/dV$ line-cuts reveal a sizable bias asymmetry.  This breakdown of $dI/dV$ particle-hole symmetry can be attributed to the bias dependence of the tunnel barrier transparencies~\cite{melo2020conductance}, an effect that does not enter the LDOS, which is symmetric by construction (Fig. 4E and Fig. 5, G and H).

As mentioned, the ZBPs in Fig. 5, A to C, are not preceded by Zeeman-split ABSs, a finding that is seemingly at odds with our previous singlet-doublet transition picture. Moreover, the naive estimation that magnetic field mediated zero-energy crossings should occur near $V_Z\approx\Delta$, because $|\zeta(B=0)|\sim\Delta$, does not explain the data, which shows ZBPs at much smaller Zeeman energies.

To clarify this, we again use our generalized superconducting Anderson model, but now in a different regime where $|\zeta(B=0)|\sim\Delta$ and with a singlet GS close enough in parameter space to the singlet-doublet boundary.  
Even in the constant pairing case, i. e., neglecting the LP modulation of $\Delta$, the magnetic field evolution of the ABSs strongly differs from that corresponding to a purely Zeeman-driven regime. This is a consequence of the non-zero coupling $\Gamma_S$ to the SC. Figure 5G illustrates this fact (compare black and yellow curves).

This can be understood from the analytical expression that governs zero-energy parity crossings
\begin{eqnarray}
\label{YSR-crossing}
\tilde V_Z=\pm\sqrt{\tilde\epsilon^2+\tilde\Gamma^2_S},
\end{eqnarray}
where $\tilde V_Z\equiv V_Z+\frac{U}{2}(\langle n_\uparrow \rangle-\langle n_\downarrow \rangle)$, $\tilde \epsilon\equiv \epsilon_0+\frac{U}{2}(\langle n_\uparrow \rangle+\langle n_\downarrow \rangle)$ and $\tilde\Gamma_S\equiv \Gamma_S+U \langle d_\uparrow d_\downarrow \rangle$. Here  
$\langle n_{\sigma} \rangle=\langle d^\dagger_\sigma d_\sigma \rangle$ and $ \langle d_\uparrow d_\downarrow \rangle$ are the QD spin-polarized occupations and the anomalous average, respectively. $\tilde V_Z$, $\tilde \epsilon$ and $\tilde\Gamma_S$ can be physically interpreted, respectively, as an exchange field, a level shift and an effective coupling that are renormalized owing to correlations [for a full derivation see \cite{supplement,ramon2015}]. Note that, for large $U$, the finite spin polarization $\langle n_\uparrow \rangle\neq\langle n_\downarrow \rangle$ that is induced at the singlet-doublet crossing explains the deviation from the pure Zeeman regime governed by $V_Z$. The same happens for large $\Gamma_S$. It is also notable that the criterion for parity crossings of YSR states in Eq. (\ref{YSR-crossing}) is essentially the same as the criterion for topological superconductivity in a proximitized NW by just substituting the renormalized QD parameters by the equivalent ones in a NW, i.e., the external Zeeman field, the chemical potential and the induced superconducting pairing:  $\tilde V_Z \rightarrow V^{NW}_Z$, $\tilde\epsilon \rightarrow\mu^{NW}$ and $\Gamma_S\rightarrow \Delta^{NW}$ \cite{lutchyn_majorana_2010,oreg_helical_2010}. This is yet another example that emphasises the difficulty of making a clear distinction between parity crossings in QDs and MZMs in NWs.
The LP modulation of the gap adds further complexity to the problem (because the ratio $\Delta(B)/U$ evolves with magnetic field), which results in a singlet-doublet transition in the first lobe at $B = 105 \, \si{mT}$ (black curves in Fig. 5H) with little resemblance to the original Zeeman-split lines (yellow lines in Fig. 5G). Together with the tunneling broadening $\Gamma_N$, this can result in a robust zero bias anomaly in the LDOS across the first LP lobe (Fig. 5H). Our theoretical analysis shows that the observed ZBPs  are actually the result of a magnetic field evolution (the full phase diagrams at three different $B$-fields are shown in Fig. 5I).

Indeed, by further investigating the gate voltage dependence of the observed ZBPs, it can be demonstrated that they do not originate from MZMs. We discuss here the behavior of devices C and E; that of device D is further presented in \cite{supplement}. We first focus on the subgap spectrum for a large range of gate voltages at zero magnetic field.  Figure 6, A and D, demonstrate that the gap is populated by ABSs. In addition, by measuring the conductance in the middle of the first lobe and sweeping the backgate voltage, one can observe that the ZBP exists only for a certain gate voltage range (Fig. 6, B and E). This is further highlighted in Fig. 6, C and F. For device C the gate range over which the ZBP persists can exceed 100 mV. However, as can be seen in Fig. 6C, the ZBP is actually the result of two ABSs merging together and splitting again for voltages around -1.8V.

\section*{Conclusions}

In summary, tunneling spectroscopy measurements on hybrid full-shell InAs/Al NWs have shown that, for short-junction devices with $X\lesssim 100 \, \si{nm}$, no ABSs or other subgap states are observed. For long junctions, a rich spectral structure arises in the LP lobes and destructive regimes. When the GS is odd, we demonstrate that the subgap excitations of the system are YSR singlets. In the metallic state within the destructive LP regions, these YSR singlets fully develop a Kondo effect, confirming our interpretation in terms of QDs. When the gap reopens, subgap YSR singlets reemerge. 
Conversely, when the GS is a singlet the flux may induce a QPT to a spin-polarized odd GS. This zero-energy fermionic parity crossing leads to a ZBP. Depending on gate conditions, this ZBP can persist for an extended magnetic-field range in the first LP lobe around $\phi\sim{\phi}_0$. When the ABS energy at zero magnetic field is close to the superconducting gap, such robust ZBPs could be mistaken for topological MZMs. The reported measurements demonstrate that, in this system, it is experimentally possible to distinguish ABSs, and their physical character, from MZMs. The competition of various physical phenomena in the same device can produce different QPTs, giving rise to ZBPs in transport spectroscopy in completely different QD parameter regimes. In the future, the reproducible fabrication of devices in which the tunnel junction does not host a QD will allow a controlled and systematic characterization of truly topological ZBPs. This is a critical step to start unveiling and exploiting the exotic physics governing topological superconductivity.

\twocolumn

\bibliographystyle{Science}

\section*{Acknowledgments}
The authors thank A. Higginbotham, E. J. H. Lee and F. R. Martins for helpful discussions. {\textbf{Funding}}. This research was supported by the Scientific Service Units of IST Austria through resources provided by the MIBA Machine Shop and the nanofabrication facility; the NOMIS Foundation and Microsoft; the European Union’s Horizon 2020 research and innovation program under the Marie Sklodowska-Curie grant agreement 844511; the FETOPEN grant agreement 828948; the European Research Commission through grant agreement HEMs-DAM 716655;  the Spanish Ministry of Science and Innovation through Grants PGC2018-097018-B-I00, PCI2018-093026, FIS2016-80434-P (AEI/FEDER, EU), RYC-2011-09345 (Ram\'on y Cajal Programme), and the Mar\'ia de Maeztu Programme for Units of Excellence in R\&D (CEX2018-000805-M); 
and the CSIC Research Platform on Quantum Technologies PTI-001. {\textbf{Author contribution}}. M.V. and G.K. designed the experiment. M.V. fabricated the devices, performed the measurements and analyzed the data under the supervision of G.K. M.B. and R.H. contributed to the device fabrication. P.K. developed the NW materials. M.V., G.K., A.H., F.P., P. S.-J., E.P. and R.A. discussed the data and contributed to their interpretation. F. P. performed the numerical simulations, with input from P. S.-J., E.P. and R.A. The analytics and interpretation of the YSR limit were developed by R. A. All authors contributed to the writing of the manuscript. {\textbf{Competing
interests}}: The authors declare no competing interests. {\textbf{Data and Materials Availability}}: All experimental data included in this work and related to this work but not explicitly shown in the paper will be available via the IST Austria repository \cite{raw_data}. Data and plot scripts from the theoretical analysis can be found online \cite{code}.

\end{document}


\baselineskip15pt


\maketitle 

\clearpage

\section*{Supplementary text}

\subsection*{Model}
\label{sec:model}
The theoretical model used to describe the experimental results of this work is presented in what follows. It consists of a single impurity Anderson model (SIAM) coupled to a superconducting reservoir, the so-called superconducting Anderson model, in which we take into account the Little-Parks (LP) modulation of the gap with the applied flux, $\phi$ (see the details below). The impurity is also coupled to a normal reservoir, hence forming a superconductor-quantum dot-normal (SC-QD-N) junction. 

The formation of the QD is taken here as an empirical fact derived from our experimental results. Its model is kept minimal, given that the precise mechanism behind its formation remains relatively unclear. Similarly, we do not add an explicit spin-orbit coupling term. Such a term would be crucial in a topological phase study, but not in the trivial regime of interest here. If the typical dot lengthscale becomes larger than the spin-orbit length (estimated in these wires to be in the hundreds of nanometers), it would modify the energy distribution and spatial profiles of the dot states, and also their couplings to the SC and N leads. As our low-energy model has a single dot orbital, any spin-orbit coupling in a more microscopic model can be understood to be already incorporated in the energy and couplings of the single dot level, which are fitting parameters in our model.

Despite its simplicity, this model correctly captures the interplay between two competing mechanisms, namely: Coulomb blockade in the QD region and Andreev processes at the superconducting interface. Each of them favors a different ground state (GS) configuration: the former favors a doublet phase, $|D\rangle$,  with spin $\frac{1}{2}$, whereas the latter a singlet phase, $|S\rangle$, with spin 0.\\
 \\ 
The Hamiltonian of the system reads: $H = H_{QD} +  H_N +H_S + H_T^N + H_T^S $. Here, $H_{QD}$ refers to the decoupled dot Hamiltonian given by
\begin{equation} 
H_{QD} =  \sum_\sigma \epsilon_\sigma d^\dagger_\sigma d_\sigma + U n_\uparrow n_\downarrow,
\end{equation} 
where $d_\sigma$ ($d_\sigma^\dagger$) destroys (creates) an electron with spin $\sigma=\{ \uparrow, \downarrow \}$ and energy $\epsilon_\sigma = \epsilon_0 \mp V_Z$, where $V_Z=g\mu_BB/2$ refers to the Zeeman energy owing to the external magnetic field $B$ (with $g$ and $\mu_B$ being, respectively, the gyromagnetic factor and the Bohr's magneton). Electron-electron correlations in the dot occupations $n_\sigma = d^\dagger_\sigma d_\sigma $ are included by means of the charging energy $U$. The Hamiltonian of the normal reservoir is given by
\begin{equation}
H_{N} =  \sum_{ k_N \sigma}  \varepsilon_{k_N\sigma} c^\dagger_{k_N \sigma} c_{k_N \sigma}, 
\end{equation}
where $c^\dagger_{k_N\sigma}$ ($c_{k_N\sigma}$) creates (destroys) an electron with momentum $k_N$, spin $\sigma$ and  single-particle energy $\varepsilon_{k_N\sigma}$  in the normal reservoir.

\begin{figure} [b!]
  \centering 
  \includegraphics[scale=.9]{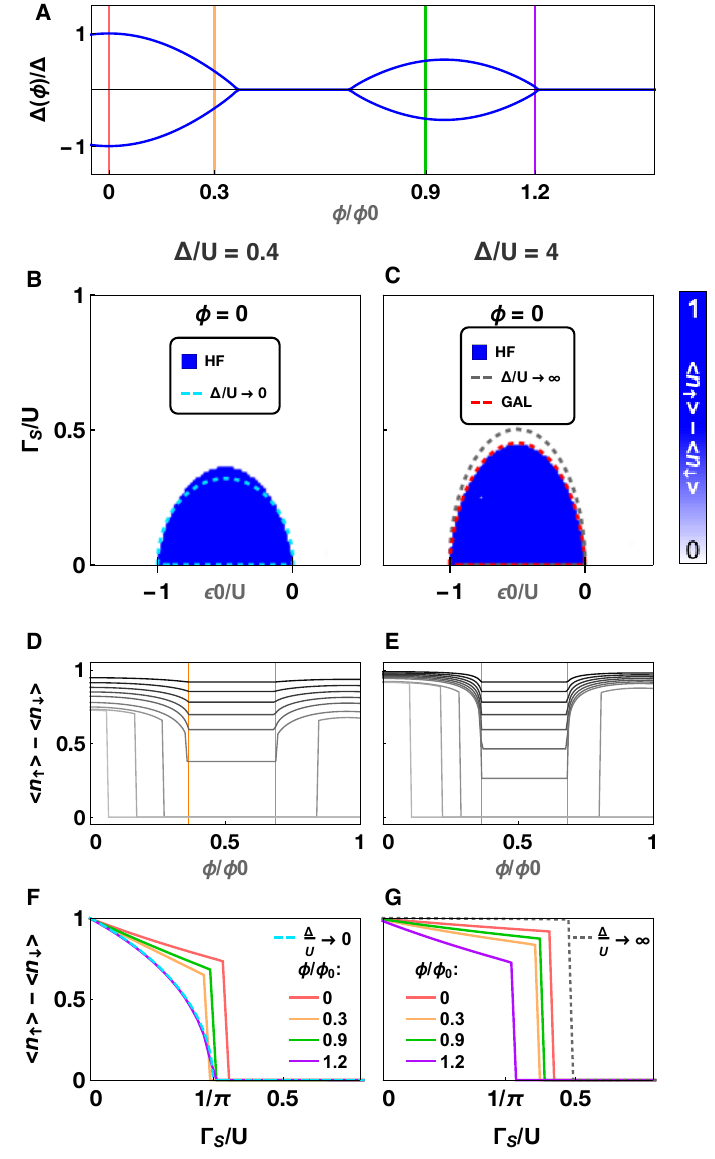}
  \caption{\label{fig:SI1} {\bf Gap modulation of the phase diagram through the Little-Parks effect}. \small{(\textbf{A}) LP modulation of the superconducting gap in the destructive regime vs normalized flux $\phi/\phi_0$ . (\textbf{B} and \textbf{C}) Full phase diagram of  
$\langle n_\uparrow \rangle- \langle n_\downarrow \rangle$ vs $\Gamma_S/U$ and $\epsilon_0/U$ for $\phi=0$ in the small and large $\Delta/U$ regimes, respectively. Only the LP effect is included here so we take $g=0$. Hartree-Fock (HF) solution: Inside the dark blue dome the superconductor-quantum dot (SC-QD) system is in a doublet ground state (GS), whereas it is in a singlet GS in the white region. Grey, red and cyan dashed lines mark the analytic boundaries in the $\Delta/U\rightarrow \infty$ (atomic), generalized atomic (GAL) and  $\Delta/U\rightarrow 0$ (Anderson) approximations, respectively. (\textbf{D} and \textbf{E}) $\langle n_\uparrow \rangle- \langle n_\downarrow \rangle$ vs $\phi/\phi_0$ for fixed $\epsilon_0=-U/2$ and increasing values of $\Gamma_S/U$ from 0.06 (black) to 0.36 (D) and 0.45 (E) (light grey). The orange lines mark the destructive LP regime. Note that for the larger couplings, the singlet phase occurs before the LP destructive region. Similarly, the system is back to a doublet well outside the LP destructive region. (\textbf{F} and \textbf{G}) $\langle n_\uparrow \rangle- \langle n_\downarrow \rangle$ vs $\Gamma_S/U$ for fixed $\epsilon_0=-U/2$ and increasing fluxes marked with colors in (A) for the two regimes of (B) and (C), respectively, also in the absence of Zeeman effect.  The LP effect results in an oscillatory modulation of the dome. Parameters: $g = 0$, $\Gamma_N \approx 0$, $\Delta=0.2$ meV, $R=50$ nm, $d=35$ nm, and $U=0.5$ meV ($U=0.05$ meV) in (B), (D) and (F) [in (C), (E) and (G)]. }}
\end{figure}

Since the superconducting lead (full-shell proximitized NW) to which the dot is coupled has a doubly connected geometry and is longitudinally threaded by a magnetic flux, $\phi=\pi R^2B$ (with $R$ being the NW radius), the LP effect will manifest in the form of a modulation of the Bardeen--Cooper--Schrieffer (BCS) gap, $\Delta(\phi)$ ~\cite{vaitieknas2019anomalous,schwiete_2010,shah_2007}. Such modulation may result in a full suppression of superconductivity in finite flux regions for cylinders characterized by small radius and thickness $d$, as compared to the zero temperature superconducting coherence length $\xi$, namely $d,R\ll\xi$~\cite{schwiete_2010}, which is the relevant experimental situation. Other than these finite size effects in the transversal direction, the finite length of the proximitized wire does not play a role since, in the experiment, this SC segment is effectively grounded. In the opposite case of a floating segment, a finite charging energy in the SC itself must be included, which significantly modifies the YSR picture presented here \cite{Pavesic:A21,Saldana:A21}. Therefore, the resulting Hamiltonian for the SC is a standard BCS model that incorporates the LP modulation of the gap and reads
\begin{equation} 
H_{S} =  \sum_{ k_S \sigma}  \varepsilon_{k_S\sigma} c^\dagger_{k_S \sigma} c_{k_S \sigma} + \sum_{k_S} \Delta(\phi) (c^\dagger_{k_S \uparrow} c^\dagger_{-k_S \downarrow} + h.c.),
\end{equation} 
where $c^\dagger_{k_S\sigma}$ ($c_{k_S\sigma}$) denotes the creation (destruction) operator in the superconducting lead with momentum $k_S$, spin $\sigma$ and energy $\varepsilon_{k_S\sigma}$. All energies are measured with respect to the chemical potential of the SC.

As was shown in~\cite{schwiete_2010}, a complete analogy can be established between pair-breaking mechanisms due to flux-induced circulating supercurrents in the shell and that of paramagnetic impurities in a disordered SC. Thus, the time-dependent Ginzburg-Landau formalism used to describe the latter can be applied to the former, giving rise to the following expression for the critical temperature $T_C (\phi)$ (see~\cite{larkin} for a comprehensive derivation):
\begin{eqnarray}
0 = \ln \left( \frac{T_C (\phi)}{T_C^0} \right) + \psi \left( \frac{1}{2} + \frac{\Lambda_n(\phi) }{2\pi T_C({\phi})} \right) - \psi \left( \frac{1}{2} \right),
\end{eqnarray}
where $T_C^0$ is the critical temperature at zero flux, $n$ is the fluxoid winding number, $\psi$ is the digamma function, and $\Lambda_n$ is the pair-breaking term corresponding to a hollow superconducting cylinder. This pair-breaking term can be derived solving the Ginzburg-Landau equations in the presence of impurities \cite{Groff:PR}, and yields
\begin{equation}
\label{Lambdanphi}
\Lambda_n(\phi) = \frac{T_C^0}{\pi} \frac{\xi^2}{(R+d)^2}\Big[ 4\left(n-\frac{\phi}{\phi_0}\right)^2 +\frac{d^2}{(R+d)^2} \left(\frac{\phi^2}{\phi_0^2} + \left( \frac{1}{3} + \frac{d^2}{20(R+d)^2}\right) n^2\right)  \Big],
\end{equation}
where $\phi_0$ denotes the superconducting flux quantum, $\phi_0=h/2e$. Once $T_C(\phi)$ is known, in small devices ($d,R\ll\xi$) the field-dependent critical current $I_C(\phi)$ can be expressed as $I_C(\phi) \propto T_C(\phi)^{3/2}$, thanks to a Ginzburg-Landau approximated relation in the presence of a magnetic field ~\cite{Bardeen1962}. In the ballistic regime, $\Delta(\phi)$ is proportional to $I_C(\phi)$ and thus
\begin{equation}
\frac{\Delta (\phi) }{\Delta} \approx \left(\frac{T_C(\phi)}{T_{C}^0} \right)^{3/2},
\label{nonlinear}
\end{equation}
where $\Delta$ is the gap at $\phi=0$.

Departures from a power-law dependence of $\Delta(\phi)$ on $T_C(\phi)$ in the presence of pair-breaking mechanisms are predicted to happen as we go beyond the Ginzburg-Landau/BCS description~\cite{abrikosov1961,gennes66,Skalski64}. However, Eq.~\eqref{nonlinear} suffices to accurately capture the LP modulation found in the experiment, as we describe in the Materials and Methods subsection.
%\begin{figure}
%\includegraphics[scale=.9]{SIdeltaT.pdf}%
%\caption{\label{fig:figureSth1}Numerical $dI/dV$ as a function of the bias voltage V and magnetic flux $\phi/\phi_0$  with the same parameters as in Fig. 4g1) (see Table~\ref{params}). In a) the BCS linear dependence of Eq.~\ref{linear} between $\Delta (\phi)$ with $T_C(\phi)$ was assumed, whilst in b) it is given by the GL result of Eq.~\ref{nonlinear}. }
%\end{figure}

The remaining two terms in the Hamiltonian:
\begin{equation} 
H_{T}^S =  \sum_{k_S \sigma} \left(V_{k_S} c^\dagger_{k_S \sigma} d_\sigma + h.c. \right),
\end{equation} 
and
\begin{equation} 
H_{T}^N =  \sum_{k_N \sigma} \left(V_{k_N} c^\dagger_{k_N \sigma} d_\sigma + h.c. \right),
\end{equation} 
stand for the couplings between the dot and the SC lead and the normal reservoir, respectively. These two tunneling processes define the two tunneling rates $\Gamma_{N,S}/\hbar$, where $\Gamma_{N,S} = \pi \sum_{k_{N,S} }|V_{k_{N,S}}|^2 \delta(\omega - \epsilon_{k_{N,S}})=\pi|V_{{N,S}}|^2\rho_{{N,S}}$ are energy broadenings, with $\rho_{{N,S}}$ being the density of states of the normal/superconducting lead evaluated at the Fermi energy.

In order to compute spectral observables, the relevant quantity that has to be determined is the equilibrium retarded Green's function in the dot, namely $\hat {\mathcal G}_\sigma(\omega)$. In a Nambu $2\times 2$ basis spanned by $\Psi_\sigma = \{d_\sigma,d^\dagger_{-\sigma}\}^T$, the Dyson equation reads
\begin{equation}
 \hat {\mathcal G}_\sigma^{-1}(\omega,\phi) =  \hat {\mathcal G}_{0 \sigma}^{-1}(\omega,\phi) - \hat{\Sigma}_\sigma (\omega,\phi).
\end{equation}
Here,  $\hat {\mathcal G}_{0 \sigma}(\omega)$ refers to the decoupled and non-interacting dot retarded Green's function and $ \hat{\Sigma}_\sigma = \hat{\Sigma}_\sigma^{QD} +\hat{\Sigma}_\sigma^N +\hat{\Sigma}_\sigma^S$ is the self-energy that takes into account the coupling to the leads, $\hat{\Sigma}_\sigma^{N,S}$, and the Coulomb interaction in the dot, $\hat{\Sigma}_\sigma^{QD} $. While a complete description of the problem needs to fully take into account electron-electron correlations,  the Coulomb blockade regime is well captured by considering a diagrammatic expansion of the dot self-energy up to the lowest order in $U$: 
\begin{equation}
\hat {\Sigma}^{QD,HF}_\sigma = U \begin{pmatrix}
\langle n_{-\sigma}  \rangle & \langle d_{\uparrow} d_{\downarrow}  \rangle\\
\langle  d^\dagger_{\downarrow} d^\dagger_{\uparrow}  \rangle & -\langle n_{\sigma} \rangle
\end{pmatrix}.
 \label{selfenergy}
\end{equation}
This mean-field approximation, which neglects high-order correlations, is known as the Hartree-Fock-Bogoliubov (HF) approximation~\cite{PhysRevB.68.035105,Rodero_2012,Lee:NN14}. This level of approximation is enough to achieve an excellent  agreement with the experimental results (away from the Kondo regime). Within this mean field picture, the retarded dot Green's function reads

\begin{equation}
\hat{\mathcal G}_\sigma^{r,HF}(\omega,\phi)= \frac{1}{D(\omega,\phi)} \small{
  \begin{pmatrix}
    \omega +\epsilon_{-\sigma} +i \Gamma_N + \Gamma_S \frac{\omega}{\sqrt{\Delta(\phi)^2-\omega^2}}+U\langle n_\sigma \rangle &  \Gamma_S \frac{\Delta(\phi)}{\sqrt{\Delta(\phi)^2-\omega^2}} + U \langle d_\uparrow d_\downarrow \rangle\\
     \Gamma_S \frac{\Delta(\phi)}{\sqrt{\Delta(\phi)^2-\omega^2}} + U \langle d^\dagger_\downarrow d^\dagger_\uparrow \rangle & \omega - \epsilon_\sigma + i \Gamma_N +  \Gamma_S \frac{\omega}{\sqrt{\Delta(\phi)^2-\omega^2}} - U \langle n_{-\sigma}\rangle
  \end{pmatrix},}
  \label{G}
\end{equation}
where $D(\omega,\phi)$ is given by the determinant of $[\hat{\mathcal G}_\sigma^{r,HF}]^{-1}(\omega,\phi)$. The Andreev bound states (ABSs) are given by the zeroes of the determinant and can be written in a compact form as:
\begin{eqnarray}
 \label{D}
\left[ \omega +i \Gamma_N + \Gamma_S \frac{\omega}{\sqrt{\Delta(\phi)^2-\omega^2}}\pm \tilde V_Z\right]^2=\tilde \epsilon^2+\left|\frac{\Gamma_S\Delta(\phi)}{\sqrt{\Delta(\phi)^2-\omega^2}}+\tilde\Delta\right|^2,
\end{eqnarray}
with $\tilde V_Z=V_Z+\frac{U}{2}(\langle n_\uparrow \rangle-\langle n_\downarrow \rangle)$, $\tilde \epsilon=\epsilon_0+\frac{U}{2}(\langle n_\uparrow \rangle+\langle n_\downarrow \rangle)$ and $\tilde\Delta=U \langle d_\uparrow d_\downarrow \rangle$. Physically, $\tilde V_Z$ and $\tilde \epsilon$ can be physically interpreted as an effective exchange field and a level shift due to correlations. When $\Delta\rightarrow 0$ they define two Coulomb resonances at $\tilde \epsilon\pm \tilde V_Z$ and broadened by $\Gamma_N+\Gamma_S$.
When $\tilde \epsilon=0$, these two resonances are located at $\pm (V_Z+U/2)$. In the presence of superconducting correlations, these resonances become ABSs whose location and physical character depend on the ratios of different parameters [$\Delta(\phi)/U$, $\Gamma_S/U$, etc.] that give rise to different physical regimes, as we explain below.

%\begin{eqnarray}
%D(\omega,\phi)&=&[ \omega +\epsilon_{-\sigma} +i \Gamma_N + \Gamma_S \frac{\omega}{\sqrt{\Delta(\phi)^2-\omega^2}}+U\langle n_\sigma \rangle ][\omega - \epsilon_\sigma + i \Gamma_N +  \Gamma_S \frac{\omega}{\sqrt{\Delta(\phi)^2-\omega^2}} - U \langle n_{-\sigma}\rangle]\nonumber\\
%&-&[\Gamma_S \frac{\Delta(\phi)}{\sqrt{\Delta(\phi)^2-\omega^2}} + U \langle d_\uparrow d_\downarrow \rangle][ \Gamma_S \frac{\Delta(\phi)}{\sqrt{\Delta(\phi)^2-\omega^2}} + U \langle d^\dagger_\downarrow d^\dagger_\uparrow \rangle].
%  \label{D}
%\end{eqnarray}
%The poles of the Green's function, the Andreev bound states (ABSs), are given by the zeroes of the determinant, namely $D(\omega,\phi)=0$. 
The average values concerning the dot level occupations in Eqs. \eqref{G} and \eqref{D} are determined self-consistently {in equilibrium}. Once $\hat{\mathcal G}_\sigma^{r,HF}(\omega)$ is known, we can easily calculate the {density of states (DOS)} $\rho(\omega,\phi) = -\frac{1}{\pi} {\rm{Im\,Tr}} [\mathcal G_\sigma^{r,HF}(\omega,\phi)]$, where the trace includes both spin and Nambu
degrees of freedom. % as well as its pole structure, which gives the Andreev bound states (ABSs), from the roots of $D(\omega,\phi)$. 
At small bias voltage $V$ and temperature $T$, the differential conductance $dI/dV$ in the tunneling limit provides an approximate measure of the QD DOS at energy $eV$. The sets of parameters for each simulation in the main text are collected in Table~\ref{params}.

\begin{table}
\begin{center}
  \begin{tabular}{ | l | c | c  | c | c | c | c | c | c |  c | }
    \hline
	 &  $U [\rm{meV}]$ & $\epsilon_0/U$ & $g$ & $\Delta [\rm{meV}]$ &  $\Gamma_S/U$ & $\Gamma_S/\Delta$ & $\Gamma_N/\Delta$ & $R$ [\rm{nm}] & $d$ [\rm{nm}]  \\ \hline
  Fig. 4F(left)   & 2.5 & -0.97 & 13 & 0.18 & 0.036 & 0.5 & 0.07 & 65 & 25 \\ \hline
  Fig. 4F(center)   & 2.5 & -0.95 & 13 & 0.18 & 0.036 & 0.5 & 0.07 & 65 & 25 \\ \hline
  Fig. 4F(right)   & 2.5 & -0.93 & 13 & 0.18 & 0.036 & 0.5 & 0.07 & 65 & 25 \\ \hline
  Fig. 5, G and H & 1.5 & -0.93 & 13 & 0.14 & 0.075 & 0.8 & 0.1    & 64 & 23 \\ \hline
  Fig. S4        & 1.5 & -0.91 & 14.3 & 0.18 & 0.12 & 1  & 0.16   & 50 & 35 \\ \hline
  Fig. S9D   & 2.5 & -0.5   & 13 & 0.18 & 0.18 & 2.5 & 0.3 & 65 & 25 \\  \hline
  Fig. S11D   & 1 & -0.5   & 13 & 0.18 & 0.33 & 1.83 & 0.22 & 65 & 25 \\ 
  %Fig. S10(B) & 1.5 & -0.93 & 17 & 0.13 & 0.062 &  0.72  & 0.25   & 64 & 23 \\
%  Fig. 4(g)   & 2.5   & -0.5      & 10     & 2.5    & 0.3    & 65    & 25 \\ \hline
%  Fig. 5(h1) & 2.5   & -0.97    & 10     & 0.5    & 0.07  & 65    & 25 \\ \hline
%  Fig. 5(h2) & 2.5   & -0.95    & 10     & 0.5    & 0.07  & 65    & 25 \\ \hline
%  Fig. 5(h3) & 2.5   & -0.93    & 10     & 0.5    & 0.07  & 65    & 25 \\ \hline
%  Fig. S8(b)   & 1.5   & -0.91    & 11     & 1       & 0.16  & 50    & 35 \\ \hline
%  Fig. S9(b)   & 1.5   & -0.93    & 13     & 0.72   & 0.25  & 64    & 23 \\ 
  \hline
 \end{tabular}
\end{center}
\caption{\label{params} \textbf{Parameters used in the numerical simulations of the main text.} \small{ $\Delta$ values were determined from experimental evidence. $R$ and $d$ have been extracted by using them as free parameters in Eqs. \eqref{Lambdanphi} and \eqref{nonlinear} and finding the best agreement with the experimental LP data (see Materials and Methods subsection). Going from flux $\phi$ to magnetic field $B$ we assumed a circular effective section of radius $R + 10 \rm{nm}$ which captures the finite penetration length of $B$ in the SC shell and is required to achieve experimental agreement.}}
\end{table}

While the full solution of the zeroes in Eq. \eqref{D} is needed to find the ABSs at arbitrary energy, one may find some useful analytic boundaries in different parameter regimes, as we discuss now.
\subsection*{Large $U$ approximations}
The experimental regime discussed in the main text corresponds to small $\Delta(\phi)/U$ (large-$U$ limit). In this case, singlets are Yu-Shiba-Rusinov (YSR)-like superpositions between the singly occupied state in the QD and quasiparticles below the gap in the SC. The ABS position in this limit can be much lower than the original gap and, assuming that $\tilde V_Z$ is the largest scale in the problem, namely that the system is in a doublet GS, we can find the poles of Eq. \eqref{D} as 
($\Gamma_N=0$):
\begin{eqnarray}
\label{YSR0}
\omega\approx \pm\Delta(\phi)\frac{\tilde V^2_Z-\tilde\Gamma^2_S-\tilde \epsilon^2}{\sqrt{[\tilde\Gamma^2_S+\tilde \epsilon^2-\tilde V^2_Z]^2+4(\tilde\Gamma_S-\tilde\Delta)^2\tilde V^2_Z}},
\end{eqnarray}
which is a generalization of the classical expression for YSR states in a SC \cite{Yu:APS65,Shiba:PTP68,Rusinov:SPJ69,RevModPhys.78.373} (see also Ref. \cite{PhysRevB.92.235422} for a different derivation using an effective Kondo model). Note that the poles cross zero energy when \cite{ramon2015}
\begin{eqnarray}
\label{YSR-crossing}
\tilde V^2_Z=\tilde\Gamma^2_S+\tilde \epsilon^2
\end{eqnarray}
We note that this criterion for parity crossings of YSR states in a QD is essentially the same as the criterion for topological superconductivity and Majorana zero modes in a proximitized NW (by just substituting $\tilde\Gamma_S\rightarrow \Delta^{NW}$, $\tilde \epsilon \rightarrow\mu^{NW}$ and $\tilde V_Z \rightarrow V_Z^{NW}$ \cite{lutchyn_majorana_2010,oreg_helical_2010}. This is yet another example that emphasises the difficulty of making a clear distinction between parity crossings in QDs and Majorana zero modes in NWs.
The connection between Eq. \eqref{YSR0} and the standard expression for YSR states  \cite{Yu:APS65,Shiba:PTP68,Rusinov:SPJ69,RevModPhys.78.373} is evident in the limit $\tilde \epsilon\rightarrow 0$, $\tilde\Delta\rightarrow 0$, where we recover
\begin{equation}
\label{YSR}
\omega=\pm\Delta(\phi)\frac{1-\left(\frac{\Gamma_S}{\tilde V_Z}\right)^2}{1+\left(\frac{\Gamma_S}{\tilde V_Z}\right)^2}=\pm\Delta(\phi)\frac{1-J^2}{1+J^2}.
\end{equation}
Using $\langle n_\uparrow \rangle-\langle n_\downarrow \rangle=1$, the exchange constant is $J=2\Gamma_S/(U+2V_Z)$, and we recover the standard expression with a exchange constant in a YSR/Kondo language $J=\pi j S\rho_S/2$, with $S=1/2$ and $j=8|V_S|^2/(U+2V_Z)$. 
%At finite Zeeman fields, we just substitute $\tilde U\rightarrow (U+V_Z)/2$ in Eqs. \ref{YSR0} and \ref{YSR} to obtain:
%\begin{equation}
%\label{YSR-Zeeman}
%\omega=\pm\Delta(\phi)\frac{(U+V_Z)^2-4\Gamma^2_S}{(U+V_Z)^2+4\Gamma^2_S}.
%\end{equation}

Typically, $U$ can be experimentally extracted from the size of Coulomb blockade diamonds and we can use Eq. \eqref{YSR} to find the value of $\Gamma_S$  in the doublet regime by just substituting the ABS energy position at $B=0$, namely $\omega=\pm \zeta$:
\begin{equation}
\label{Gamma_s}
\Gamma_S=\frac{U}{2}\sqrt{\frac{(1-\zeta/\Delta)}{(1+\zeta/\Delta)}}.
\end{equation}
Using this procedure, we find a good agreement between the analytical YSR expression of Eq. \eqref{YSR}  and the experiments in the full magnetic field range [see Fig. 3D of the main text]. Further improvement in the agreement can be found by performing full HF numerics directly from Eq. \eqref{D} [typically using slightly smaller $\Gamma_S$ values as the ones directly extracted from Eq. \eqref{Gamma_s}]. Note that the way the magnetic field enters in Eqs. \eqref{YSR0} and \eqref{YSR} (through $\Delta(\phi)$ and $V_Z$) strongly deviates from a standard linear Zeeman dispersion, which explains why the ABSs in the experiments show almost no dispersion against magnetic field.

These YSR singlets eventually become Kondo singlets in the destructive LP regions when $\Delta(\phi)\rightarrow 0$. One can estimate the flux at which Kondo physics dominates over superconductivity by using the expression for the $V_Z=0$ Kondo temperature at the symmetry point $T_K\approx0.3\sqrt{\Gamma_SU}e^{-\frac{\pi U}{8\Gamma_S}}\approx0.2U\sqrt{J}e^{-\frac{\pi}{4J}}$.  % (the precise prefactor at $\epsilon_0\neq-U/2$ is not well defined). 
For example, using the parameters of Fig. 4(g) of the main text, we can estimate that a Kondo GS develops when $\Delta(\phi)\lesssim T_K\approx 0.044\; \rm{meV}\approx 0.24\Delta$, which first happens at $\phi\approx 0.35\phi_0$, in very good agreement with the experimental observation of a Kondo effect in the first destructive LP region at $B\approx50$ mT. Nevertheless, we emphasise that a full calculation of the flux at which the Kondo effect develops constitutes a rather involved problem, well beyond the scope of this theoretical analysis, and should be the subject of future studies.

When approaching a singlet-doublet crossing near a charge-degeneracy point as a function of $\epsilon_0$, the self-consistent HF method is known to exhibit small discontinuities between non-magnetic and magnetic solutions~\cite{Rodero_2012,Lee:NN14}. In these cases (Fig. 4E of the main text), and since these discontinuities can result in unphysical energy shifts of the ABSs near the crossing, we find the mean-field values in Eq. \eqref{D} that provide the best agreement with the experiment.

%Within this approximation the average values concerning the dot level occupations have to be determined self-consistently. However, arguing that self-consistency will result in small quantitative (and not qualitative) changes, we fix the average values in \eqref{selfenergy} as those corresponding to the large gap $\Delta \rightarrow \infty$ limit in which the self-energy becomes static.  %In this limit, the ground state ($|D\rangle$ or $|S\rangle$) of the system is fixed by the mean field transtition line of the superconducting Anderson model in the presence of a Zeeman field, which is given by: %$(V_Z + \frac{U}{2})^2 = \Gamma_S^2 + (\epsilon_0 + \frac{U}{2})^2$, leading to a simple analytical formula for $\hat{\mathcal G}_\sigma^{r,HF}(\omega)$ . %Once $\hat{\mathcal G}_\sigma^{r,HF}(\omega)$ is known, we can easily calculate the spectral function  using: $ A(\omega,\phi) = -\frac{1}{\pi} ImTr [\mathcal G_\sigma^{r,HF}(\omega,\phi)]$ which is proportional to the differential conductance $dI/dV$ in the tunneling limit, as well as its pole structure (Andreev bound states) from the roots of $D(\omega,\phi)$. The set of parameters for each simulation is collected in Table~\ref{params}. 

\subsection*{Large $\Delta$ approximations}
For completeness, we also discuss the opposite large gap limit, where the coupling to the SC $\Gamma_S$ mainly induces local superconducting correlations in the QD. This can be easily seen by taking the so-called $\Delta\rightarrow\infty$ atomic limit \cite{ramon2015,Bauer_2007,PhysRevB.79.224521}, where superconducting singlets are just bonding and antibonding superpositions of empty and doubly-occupied states in the QD, namely $|S_\pm\rangle=|0\rangle\pm|2\rangle$. In this atomic limit, the singlet-doublet boundary (parity crossing) can be easily obtained by taking the $\omega\rightarrow 0$ limit in Eq. \eqref {D} ($V_Z=0$ and $\Gamma_N=0$):
\begin{equation}
\left(\frac{U}{2}\right)^2 = \Gamma_S^2 + \left(\epsilon_0 + \frac{U}{2}\right)^2.
\end{equation}
Expanding the determinant further to order $\omega^2$ allows to capture finite $\Delta$ corrections to the above atomic limit boundary. This is called the generalized atomic limit (GAL) \cite{GAL1,GAL2}, whose singlet-doublet boundary now reads 
\begin{equation}
\frac{U^2}{4(1+\frac{\Gamma_S}{\Delta})^2} = \Gamma_S^2 + \left(\epsilon_0 + \frac{U}{2}\right)^2.
\end{equation}
It can be used to demonstrate that HF provides a very good approximate solution of the problem, particularly near the symmetry point $\epsilon_0=-U/2$, when neglecting the above-gap continuum contribution to the anomalous propagator.
Further improvement can be reached by performing finite-$\Delta$ self-consistent perturbation theory around the atomic limit both for $V_Z=0$ \cite{PhysRevB.79.224521} and $V_Z\neq 0$ \cite{PhysRevB.94.085151,PhysRevResearch.2.012065}. The resulting boundaries are, however, no longer analytic. 

\begin{figure}[t!]
\includegraphics[scale=.9]{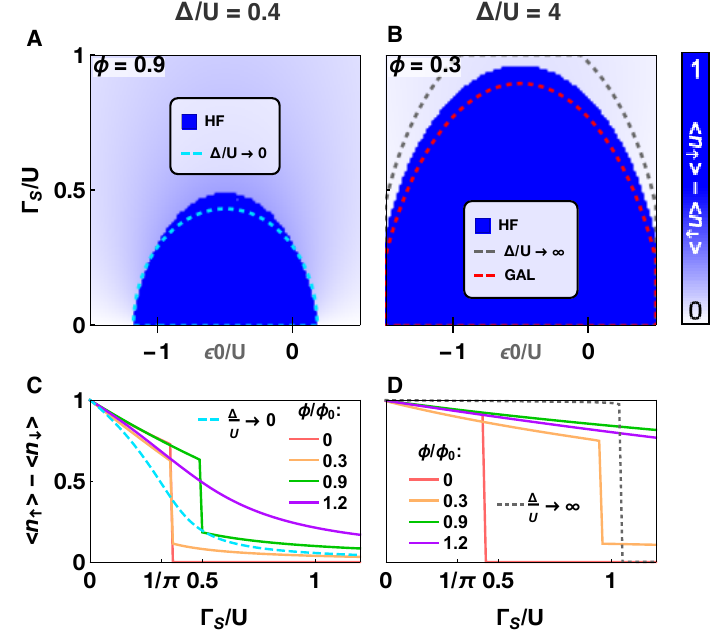}%
\caption{\label{fig:SI2} {\bf Gap plus Zeeman modulation of the phase diagram owing to external flux}. \small{ (\textbf{A} and \textbf{B}) Full phase diagram of  $\langle n_\uparrow \rangle- \langle n_\downarrow \rangle$ vs $\Gamma_S/U$ and $\epsilon_0/U$ at $\phi/\phi_0= 0.9$ ($\phi/\phi_0= 0.3$) in the small (large) $\Delta/U$ regime. Hartree-Fock (HF) solution: Inside the dark blue dome the SC-QD system is in a doublet ground state (GS), whereas it is in a singlet GS in the white region. Grey, red and cyan dashed lines mark the analytic boundaries in the $\Delta/U\rightarrow \infty$ (atomic), generalized atomic (GAL) and  $\Delta/U\rightarrow 0$ (Anderson) approximations, respectively. (\textbf{C} and \textbf{D}) $\langle n_\uparrow \rangle- \langle n_\downarrow \rangle$ vs $\Gamma_S/U$ for fixed $\epsilon_0=-U/2$ in the small and large $\Delta/U$ regimes, respectively, and increasing fluxes marked with colors in (A). The dashed cyan line in (C) corresponds to the $B=0$, $\Delta/U\rightarrow 0$ limit, whereas the dashed grey line in (D) corresponds to the $B=0$, $\Delta/U\rightarrow\infty$ limit. In this figure, both the Zeeman effect, through a non-zero $g = 11$, and the LP effect are included, the rest of parameters are chosen as those of Fig.~\ref{fig:SI1}. The doublet dome exhibits a non trivial evolution with the applied flux due to two different mechanisms: the Zeeman effect that expands the doublet dome in both $x$ and $y$ directions and the LP effect which leads to an oscillatory modulation.}}
\end{figure}
\subsection*{Phase diagrams in different coupling regimes}

Whether the SC-QD GS is a singlet or a doublet is revealed by the magnetic field evolution of the subgap states, with a change in GS parity whenever they cross zero energy. This in turn results from the competition of two mechanisms:  LP effect that modulates the gap and the Zeeman effect that shifts the dot level energy. The first effect is illustrated in Fig. \ref{fig:SI1}A, where we show the oscillatory modulation of the superconducting gap vs the applied normalized flux $\phi/\phi_0$. The gap exhibits the so-called destructive LP effect, whereby the gap decreases in the zeroth lobe from its maximum value at $\phi=0$, disappears in the first destructive region, reemerges in the fist lobe, and so on. In Fig. \ref{fig:SI1}, B and C, we analyze the role of LP modulation of the gap on the phase diagram of the problem, disregarding for the moment the Zeeman effect (i.e., taking $g=0$). We plot the difference of spin-polarised occupations $\langle n_\uparrow \rangle- \langle n_\downarrow \rangle$ in the self-consistent HF approximation versus $\Gamma_S/U$ and $\epsilon_0/U$. Both panels show the phase diagram for the largest gap of the problem, $\Delta$ (i.e., the gap at $\phi=0$), but for different charging $U$, resulting in small, Fig. \ref{fig:SI1}B, and large, Fig. \ref{fig:SI1}C, $\Delta/U$ ratios. Already at this level, it it important to stress how the different ratios influence the singlet-doublet boundary. The gap evolution across LP oscillations (without Zeeman) modifies the singlet-doublet boundary most strongly at the symmetry point $\epsilon_0=-U/2$. This LP modulation of the boundary is clearly seen in Fig. \ref{fig:SI1}, D and E, where we show how the spin polarisation varies with flux at the symmetry point $\epsilon_0=-U/2$ for increasing $\Gamma_S$ (from top to bottom curves). Both panels show that doublet-singlet-doublet transitions can be induced by LP modulation. Note that while our HF does not provide a quantitative result at doublet phases (the spin should be fully screened in this phase, $\langle n_\uparrow \rangle- \langle n_\downarrow \rangle\rightarrow 0$, owing to Kondo correlations), the overall trend is perfectly captured. These non-monotonic doublet-singlet-doublet transitions are, in turn, responsible for the phenomenology discussed in Fig. 4 of the main text. In this regime, the measured excitations in the experiments clearly show a YSR singlet-Kondo (co-tunneling) excitation-YSR singlet transition mediated by flux. Note that for the largest coupling $\Gamma_S$ shown in the plot, the boundaries occur well before/after the destructive LP region (delimited by orange lines). In this latter case, the singlet phase extends well beyond the destructive LP region. This could result in a Kondo effect and a singlet-doublet parity crossing in flux regions well inside the first lobe where Majoranas are also predicted.
Apart from a non-monotonic dependence of the boundary on $\phi$, owing to the non-monotonic $\Delta(\phi)$, we clearly observe how the two analytic boundaries of the problem, the normal Anderson model boundary at $\Gamma_S/U=1/\pi$ in the $\Delta\rightarrow 0$ (cyan dashed line) and the $\Delta/U\rightarrow\infty$ boundary at $\Gamma_S/U=0.5$ (grey dashed line), are approached. The first limit, in particular, is approached in the destructive LP phases where the superconducting gap fully closes. As we mentioned above, a Kondo effect (not captured here) should develop when $T_K\gtrsim \Delta(\phi)$ in these Anderson regions \cite{doi:10.1143/JPSJ.69.1812,Bauer_2007,Lee:PRB17}. 

Having discussed the non-trivial role that LP modulation has on the singlet-doublet boundary, we now take into account also the Zeeman effect. The full phenomenology showing the interplay of LP and Zeeman effects induced by $\phi$ is shown in Fig. \ref{fig:SI2}. The upper plots show the full phase diagram for small [Fig. \ref{fig:SI2}(a)] and large [Fig. \ref{fig:SI2}(b)] $\Delta/U$ ratios, respectively. The role of Zeeman is now evident with growing doublet regions as $\phi$ increases. As expected, for the same $g$-factor (which we take as $g=11$ here) the degree of spin-polarisation of the GS also depends on whether $U$ is much larger (Fig. \ref{fig:SI2}A) or comparable (Fig. \ref{fig:SI2}B) to the typical Zeeman energy scales. The latter limit, in particular, results in a greatly enhanced doublet dome even for moderate fluxes. This can be easily understood already from the $\Delta\rightarrow\infty$ atomic limit which now reads
\begin{equation}
\left(V_Z+\frac{U}{2}\right)^2 = \Gamma_S^2 + \left(\epsilon_0 + \frac{U}{2}\right)^2.
\end{equation}
The lower panels, Figs. \ref{fig:SI2}, C and D, show typical cuts for increasing flux at the symmetry point $\epsilon_0=-U/2$. Important for our discussion is that the singlet-doublet boundary for $\phi=0$ in the small $\Delta/U$ regime, Fig. \ref{fig:SI2}C, moves to considerably larger $\Gamma_S/U$ ratios (see e.g. the $\phi=0.9\phi_0$ boundary) due to the reentrant LP gap combined with the increased doublet phase owing to Zeeman. This overall increase of the doublet phase near $\phi\approx\phi_0$ greatly favours singlet-doublet parity crossings in the first LP lobe as we discuss now.

\begin{figure}[t!]
\includegraphics[scale=.9]{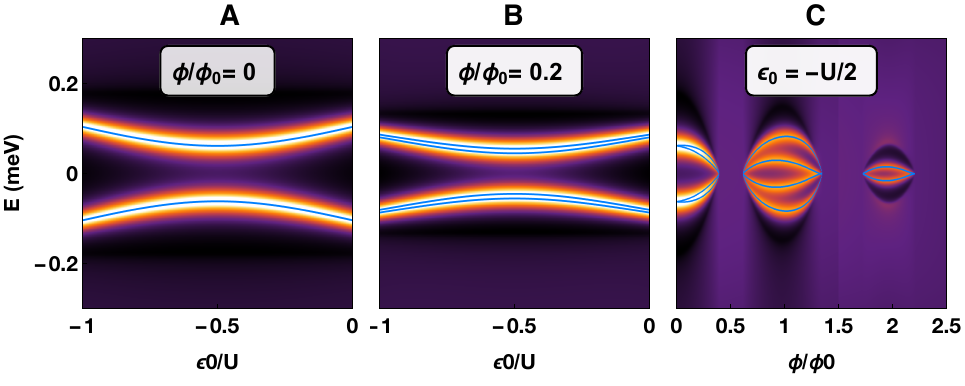}%
\caption{\label{fig:SI3} {\bf Suppressed Zeeman splitting in the zeroth lobe and ZBAs induced by singlet-doublet parity crossings.} \small{(\textbf{A} and \textbf{B}) Poles of Eq. \eqref{G} shown in blue superimposed to the numerical DOS as a function of the bias voltage $V$ and $\epsilon_0/U$ for two different normalized fluxes $\phi/\phi_0$ within the zeroth LP lobe, namely 0 and 0.2, respectively. (\textbf{C}) Poles and numerical DOS versus $V$ and $\phi/\phi_0$ at $\epsilon_0=-U/2$. $U=1$ meV, $\Gamma_S=0.43$ meV, and $\Gamma_N=0.2$ meV. The rest of parameters are the same as in Fig. 4F left panel (see Table \ref{params}). Within the zeroth lobe, the LP modulation of the gap together with a small Zeeman energy (even for a sizeable $g=13$) lead to a Zeeman splitting that cannot be experimentally resolved, as shown in (B). Using the above parameters $V_Z/\Gamma_N = 0.45$.}}
\end{figure}

\subsection*{Parity crossings versus flux $\phi$}

As it is clear from the previous discussion, the fate of the GS character with external flux depends on the initial $\phi=0$ position of the system parameters in the $\epsilon_0/U$-$\Gamma_S/U$ plane. If these parameters are far away from the singlet-doublet boundary, the system will remain in the same GS for moderate fluxes. This is the case of Figs. 4, A, B and D, of the main text where the system remains in a singlet GS for fluxes deep into the first lobe. Since the GS is a singlet, a further effect, apart from LP modulation, is the Zeeman splitting of the subgap excitations. However, such Zeeman splitting is not clearly seen in the experiment across the zeroth lobe. Figure \ref{fig:SI3} shows theoretical calculations in this regime. In agreement with the experiment, the Zeeman splitting in the zeroth lobe cannot be resolved. This can be explained by the combined effect of the LP gap closing (which strongly modifies the linear Zeeman dispersion at low $B$ fields), the small Zeeman energy (even for a sizeable $g=13$) and the tunneling broadening of the lines (compare poles with the full DOS). In Fig. \ref{fig:SI3}B $V_Z/\Gamma_N = 0.45$. 

When the initial point is relatively close to the dome boundary, one or more singlet-doublet transitions will probably occur at moderate fluxes. %In particular, the singlet-doublet transition is approached faster with flux near the center of the dome, $\epsilon_0\approx -U/2$, and for intermediate values of $\Gamma_S$. 
In particular, Fig. 5I in the main text discusses how we approach the singlet-doublet transition that we present in Figs. 5, A, B and C, (the corresponding point in parameter space for said figure is marked by a yellow circle). The system starts in a singlet GS for $\phi=0$. As the applied flux increases, the dome approaches the yellow point which crosses the singlet-doublet line at $\phi\approx\phi_0$, remaining in the doublet phase afterwards. 
We note that, if we had a system with an initial singlet state slightly different from the one of the yellow point, but same $g$-factor, it may cross the dome at very different values of  $\phi$. For example, experimental points in this phase diagram with the same $\Gamma_S$ but slightly different gate voltage outside the region $\epsilon_0/U\in [-1,0]$ (red circle), or for the same gate voltage ($\epsilon_0$) but larger coupling to the SC $\Gamma_S$ (light blue circle), are still in the singlet phase for the largest  $\phi$ shown here. This exemplifies how QD physics can lead to $\phi$-induced parity crossings, and concomitant zero bias anomalies in transport, at very different points in the parameter space of this complex interplay of different physical phenomena.

The simulation for the specific parity crossing of Fig. 5B is shown in Fig. 5, G and H, where we plot the poles of Eq. \eqref{G}, shown in black, superimposed to the numerical DOS as a function of bias voltage $V$ and magnetic flux.

\begin{figure}[t!]
\includegraphics[scale=.9]{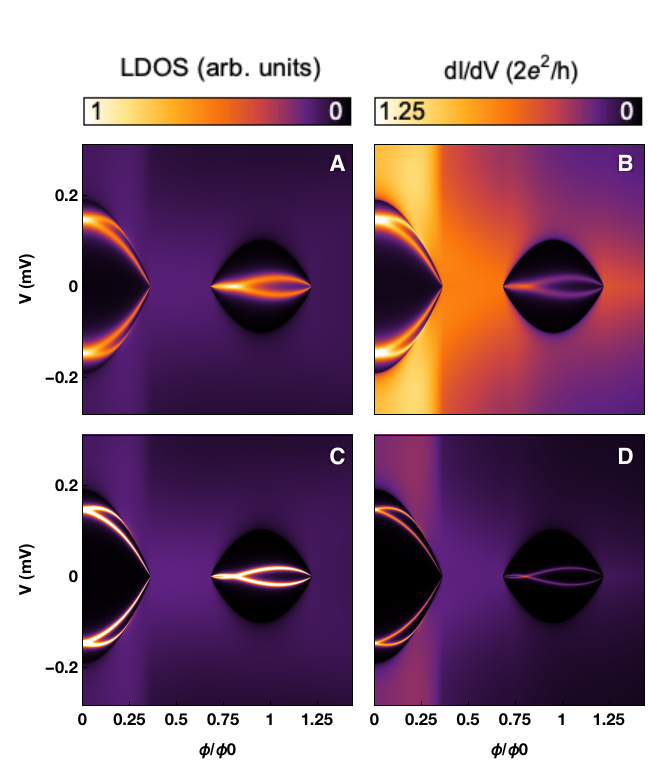}
\caption{\label{fig:dIdV}Comparison between the numerical LDOS (\textbf{A} and \textbf{C}) and the simulated $dI/dV$ (\textbf{B} and \textbf{D}) in the tunneling limit, as a function of magnetic flux $\phi/\phi_0$ in a QD-SC full shell NW. We assume a constant coupling $\Gamma_N$ to the normal probe and compare the two quantities for two small values of $\Gamma_N/\Gamma_S$ (0.16 and 0.04 in the top and bottom rows, respectively). In general, while the relative magnitude of the two quantities may differ quantitatively, the evolution of subgap ABSs are accurately captured by the peaks of the tunneling $dI/dV$ (the parameters can be found in Table \ref{params}).}
\end{figure}

\subsection*{Differential conductance versus local density of states}

A common technique to experimentally measure the spectral density of a mesoscopic system is to use a metallic probe weakly coupled to the system. Under a number of ideal assumptions (weak, energy-independent coupling $\Gamma_N$, featureless density of states in the probe with a sufficient energy range, and sufficiently low temperatures), the tunneling differential conductance $g(V) = dI/dV$ at a bias $V$ from the probe into a certain point in the system is proportional to the system's local density of states (LDOS) at energy $\epsilon = eV$. This is the basis of transport spectroscopy.

In realistic conditions, however, particularly in the case of hybrid semiconducting NWs of interest here, one can expect some deviations between the tunneling spectroscopy $g(V)$ and the Bogoliubov-de Gennes (BdG) LDOS($\epsilon$) of the coupled probe-system. In the tunneling limit, sharp peaks in the LDOS (from subgap ABSs) yield peaks in the $dI/dV$ at essentially the same energy. The LDOS peaks are delta functions of weight one, broadened phenomenologically by the N-QD coupling $\Gamma_N$. In contrast, the corresponding subgap $dI/dV$ is given by Andreev reflection peaks of height $\leq 2e^2/h$ (or $\leq 4e^2/h$ for spin-degenerate states), depending on the relative transparencies of N-QD and QD-S barriers. The width of the $dI/dV$ peaks also depends on the specific microscopic details of these barriers and may exhibit some bias dependence. Other typical deviations include bias asymmetries $g(-V)\neq g(V)$~\cite{melo2020conductance}, which are absent in the BdG LDOS by construction and, more generally, differences in the relative magnitude between subgap and above-gap values when comparing $dI/dV$ and LDOS. If we consider more realistic coupling models for the probe, across, say, an extended smooth barrier, the relative $dI/dV$ visibility of different spectral LDOS features, particularly above the gap, might change. An example of this effect relevant to Majorana wires with QDs was discussed in Ref. \cite{Prada2012}, where a smooth extended barrier between probe and system washed out small-momenta spectral features in the $dI/dV$, such as the topological gap inversion. In spite of these subtleties, tunneling spectroscopy remains a valuable technique to access the spectral density of a hybrid NW such as ours, and in particular to measure the evolution of the energy of subgap Andreev bound states, as long as $\Gamma_N$ is smaller than other relevant energy scales in the system ($\Gamma_S, k_BT, \Delta$ and $U$).

To get a more precise sense of how the $dI/dV$ approximates the LDOS in our QD-SC system, we have computed the two quantities using the same model of Section \ref{sec:model}. No further complexity is added to the probe-system coupling beyond a constant spin-independent $\Gamma_N$. The subgap $dI/dV$ (Andreev conductance) is then given by $g(V) = g_0 4\Gamma_N^2\sum_{\sigma}|G_{12,\sigma}(\omega=eV)|^2$, where $g_0=2e^2/h$ and $G_{12,\sigma}$ is the off-diagonal component of Eq. \eqref{G}. An expression for the full $dI/dV$ including above-gap quasiparticle contributions can be found in Refs.~\cite{Baranski:JPCM13,Lee:NN14}. In Fig. \ref{fig:dIdV} we show the full $dI/dV$ and the LDOS for two values of $\Gamma_N/\Gamma_S\ll 1$ (tunneling limit).
In general, the $dI/dV$ peak heights are bias dependent, and their height relative to the quasiparticle conductances above the gap depends on $\Gamma_N$. Far from the tunneling limit, i.e. $\Gamma_N/\Gamma_S\gtrsim 1$, the $dI/dV$ peaks might even acquire energy shifts for increasing $\Gamma_N$ (not shown).  In the tunneling regime of interest here, the subgap $dI/dV$ always tracks the evolution of ABSs in the LDOS as a function of parameters in a very accurate way. Thus, the only effect of a finite $\Gamma_N\ll\Gamma_S$ is merely to induce a finite broadening in the subgap peaks, both in the LDOS and the $dI/dV$ (compare the upper row with the lower row in Fig. \ref{fig:dIdV}).

Since the fine details of the probe-QD and QD-SC couplings are largely unknown in our devices, we opt in the main text to directly simulate the intrinsic LDOS of the simplified QD-SC model explained above, and compare it to the measured tunneling $dI/dV$ spectroscopy. Given the good match of our theory simulations of ABS levels to the measured evolution of $dI/dV$ peaks, we conclude that  microscopic barrier details are inessential to the interpretation of our experiment.

\section*{Additional experimental data}

\subsection*{Materials and methods}
\begin{figure}[h]
\includegraphics[scale=1]{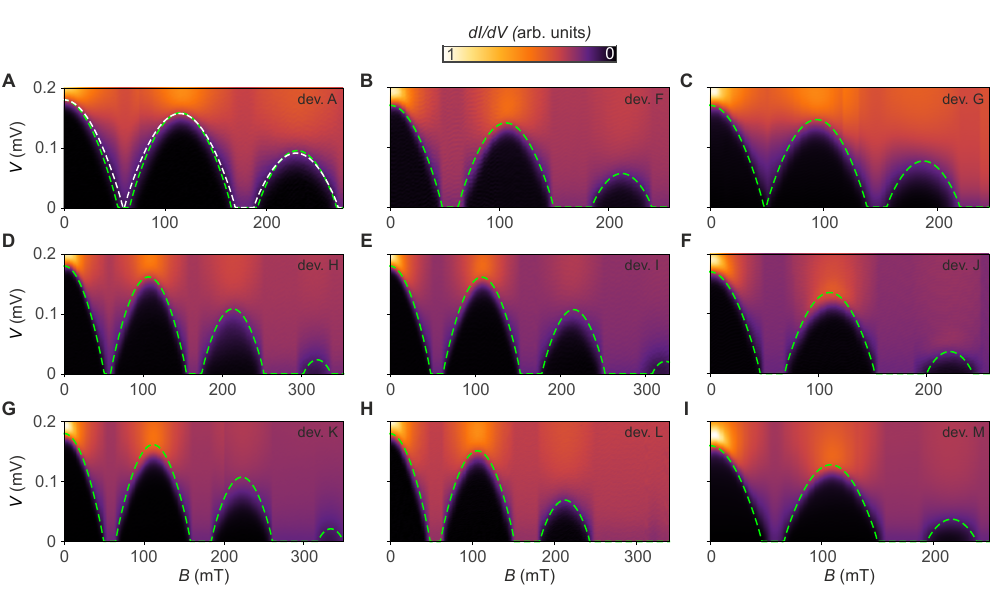}
\caption{\label{fig:figure1_part5_sup_vScience} \textbf{Little-Parks measurement used for extracting the NW parameters for all short-junction devices.} \small{ $dI/dV$ as a function of $V$ and $B$ for device A and F-M. The superimposed LP theory curves 
allow us to extract the NW parameters from the lobes using Eqs. \eqref{Lambdanphi} and \eqref{nonlinear}. For device A the white dashed curve represents the case where $R =  60 \, \si{nm}$, $d = 28 \, \si{nm}$ and $\xi  = 150 \, \si{nm}$ while the green dashed curve corresponds to $R = 64 \, \si{nm}$, $d = 24 \, \si{nm}$ and $\xi  = 160 \, \si{nm}$}. For all devices the parameters are summarized in table S2. The linear colorscale is renormalized by the maximum for each panel. }
\end{figure}

The hexagonal InAs NWs, with a diameter of about $110 \, \si{nm}$, were grown via the VLS technique and the  $30 \, \si{nm}$ Al shell was epitaxially grown in situ, giving a highly transparent semiconductor-SC interface \cite{chang_hard_2015,krogstrup_2015}. 
The NWs were deposited on a heavily doped silicon substrate covered with $285 \, \si{nm}$ of silicon oxide. The Al shell is completely removed from one side of the NW via transene etching and the bare InAs is employed as a tunnel barrier. The Al shell, $L$, is longer than $1.5 \, \si{\micro m}$ for all the measured devices. Hence, effects of overlapping potential MZMs could be neglected as $L$ is much larger than the coherence length \cite{sarma2012splitting,albrecht_exponential_2016}. The gates and  contacts consist of normal-metal ($5 \, \si{nm}$/$180 \, \si{nm}$ Ti/Au bilayer). The differential conductance, $dI/dV$, was measured via a lock-in technique in a dilution refrigerator with a base temperature of $20 \, \si{mK}$. Tunneling $dI/dV$ measurements of the superconducting gap as a function of flux are used in Fig. \ref{fig:figure1_part5_sup_vScience} and Table \ref{params2} to extract estimates for relevant NW parameters (NW radius $R$, shell thickness $d$ and coherence length $\xi$) in different devices. This is done by fitting the shape of the LP lobes against the Ginzburg-Landau theory predictions, see Eqs.  \eqref{Lambdanphi} and \eqref{nonlinear}. Note that slightly different values of the parameters produce fittings with similar level of accuracy, see white and green dotted curves in Fig. \ref{fig:figure1_part5_sup_vScience} for device A. This means that the values of these parameters can be estimated only approximately.\newline

\subsection*{Kondo effect and g-factor}

As mentioned, YSR singlets are the superconducting counterparts of Kondo singlets and a full screening of the quantum impurity spin is expected when the exchange between the QD and the SC is large enough, such that $k_BT_K\approx \Delta$.  While previous works have studied this doublet-YSR singlet quantum phase transition (QPT) by a controlled tuning of $\Gamma_S$ \cite{Lee:PRB17}, we here can reach a Kondo regime owing to the destructive LP regions where the gap is suppressed. From Eq.~(1) in the main text, we can extract an effective exchange coupling $J=2\Gamma_S/U$ (at zero magnetic field), which enters the expression of the Kondo temperature $k_BT_K\approx0.3\sqrt{U\Gamma_S} e^{-\frac{\pi U}{8\Gamma_S}}\approx 0.2U\sqrt{J}e^{-\frac{\pi}{4J}}$ at the symmetry point (the dashed line in Fig.~3A of the main text). Using the experimental values for $U$ and $\Gamma_S$, we get $k_BT_K\approx 0.044\; \rm{meV}\approx 0.24\Delta(\phi=0)$. Owing to the LP reduction of the gap, a full Kondo effect is expected to develop when $k_BT_K\approx \Delta(\phi)$ \cite{doi:10.1143/JPSJ.69.1812}, which first happens at $\phi\approx 0.35\phi_0$. Indeed, we find
a symmetric pair of excitations that appear in the first destructive LP regime,  starting at  $\approx 50$mT which corresponds to $0.33 \phi_0$ (cyan trace in Fig.~\ref{fig:figure_sup7}C), that we interpret as cotunneling processes through a spin-split Kondo resonance, supporting the previous picture. At larger fluxes, the gap reopens, and the system goes back to a doublet GS with YSR singlet  subgap excitations. This remarkable modulation from YSR singlets to Kondo and back is only possible due to the destructive LP effect. This YSR-Kondo QPT mediated by the LP effect occurs for typical QD parameters around the same flux regions where Majoranas are predicted to occur in full-shell NWs \cite{Vaitiekenaseaav3392,fern2019evenodd}, which adds further complexity to the problem.  
This Kondo interpretation is supported by the decreasing height of the conductance peaks for increasing temperatures (Fig.~\ref{fig:figure_sup7}E). 
As expected, the hard superconducting gap precludes the formation of a Kondo resonance, which is absent in the zeroth and first lobe. On the contrary, the soft gap in the second lobe allows the formation of an above-gap split Kondo resonance, which, in this regime, coexists with the YSR states \cite{Lee:PRL12,ramon2015} (see line traces in Fig.~\ref{fig:figure_sup7}C). The progression of the Kondo peaks in magnetic field allows for an estimation of the NW's $g$-factor. The extracted values are $g  = 6.4 \pm 0.8$ in the first destructive region, $g  = 9.2 \pm 0.3$ in the second destructive one 
and $g  = 10.5 \pm 0.1$ from the end of the second lobe to  $B = 900\, \si{mT}$ (Fig.~\ref{fig:figure_sup5}).  The data indicate a non linear behaviour of $g$ with the magnetic field. This apparent non-linearity relates to the LP effect as the flux which can penetrate the core of the NW depends on the magnetic field value. Nonlinearities of the effective $g$-factor owing to Kondo interactions in a superconducting system are also expected \cite{ramon2015}.

\begin{table}
\begin{center}
  \begin{tabular}{ | c| c | c  | c | }
    \hline
Name of  the device	 & $R$ [\rm{nm}] & $d$ [\rm{nm}] &  $\xi$ [\rm{nm}] \\ \hline
  A   & 64   & 24      & 160    \\ \hline
  A2   & 60   & 28      & 150    \\ \hline
  F & 60   & 28   & 165      \\ \hline
  G & 61   & 29    & 150    \\ \hline
  H & 67   & 23    & 160     \\ \hline
  I   & 66   & 21    & 165   \\ \hline
  J  & 60   & 28    & 175    \\ \hline
  K & 66   & 21    & 165     \\ \hline
  L   & 60   & 27    & 160   \\ \hline
  M  & 60   & 29    & 175     \\ 
  \hline
 \end{tabular}
  \label{params2}
\end{center}
\caption{ \label{params2} \textbf{Dimensions of short junction devices.} \small{ NW parameters extracted from the analysis of Fig. S5. Going from flux $\phi$ to magnetic field $B$ we assumed a circular effective section of radius $R + 10 \rm{nm}$ which captures the finite penetration length of $B$ in the SC shell and is required to achieve experimental agreement.}}
\end{table}

\begin{figure} [b!]
  \centering 
  \includegraphics[scale=1]{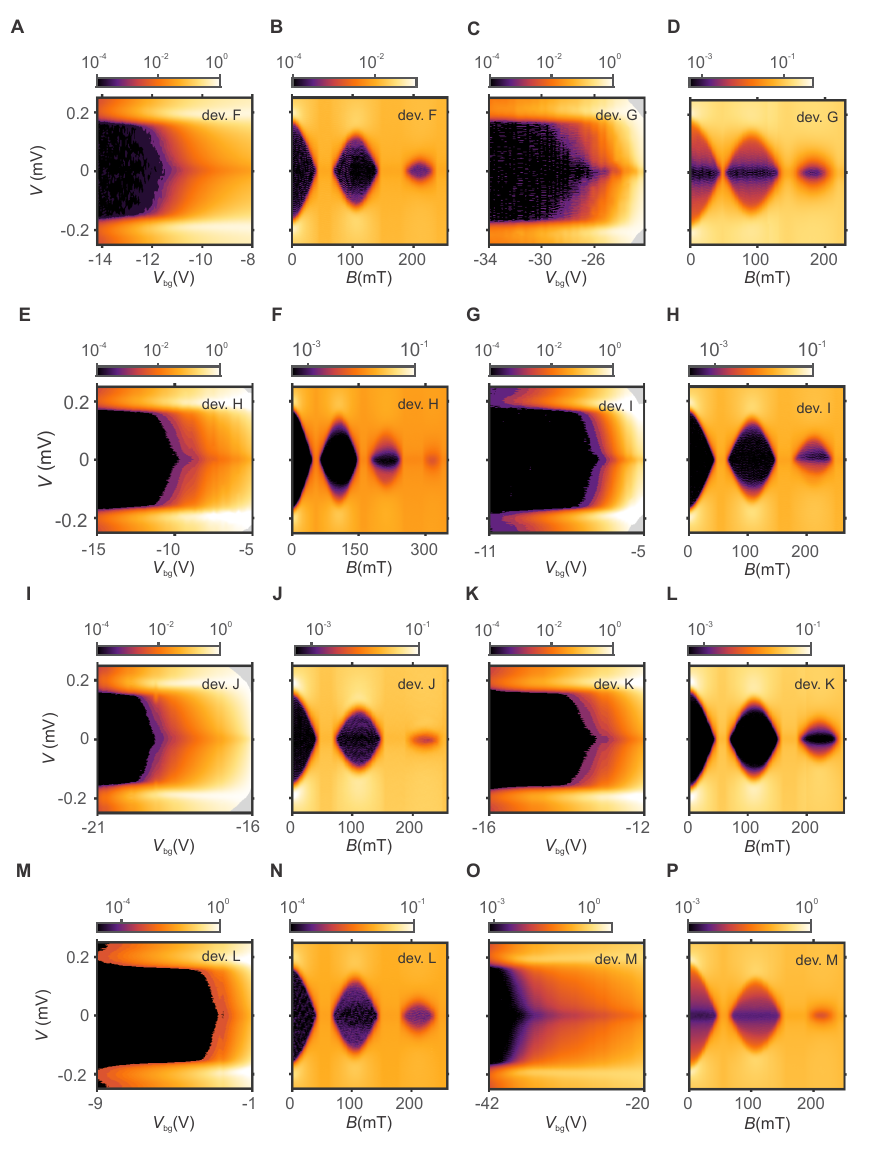}
\end{figure}
\begin{figure} [t!]
\caption{\label{fig:figure_sup1} \textbf{Devices with a short junction.} \small{ In the right-upper corner of each plot, the letter indicates the device. (\textbf{A}), (\textbf{C}), (\textbf{E}), (\textbf{G}), (\textbf{I}) , (\textbf{K}), (\textbf{M}) and (\textbf{O}) $dI/dV$ as a function of  $V$ and  $V_{\textrm{bg}}$ for devices with a short junction in absence of magnetic fields. No ABSs are present for these devices because of the short junction. More concretely, device F has $X \approx 95 \, \si{nm}$, device G has $X \approx60 \, \si{nm}$, device H has $X \approx85 \, \si{nm}$, device I has $X \approx70 \, \si{nm}$, device J has $X \approx90 \, \si{nm}$, device K has $X \approx60 \, \si{nm}$, device L has $X \approx80 \, \si{nm}$ and device M has $X \approx40 \, \si{nm}$. (\textbf{B}), (\textbf{D}), (\textbf{F}), (\textbf{H}), (\textbf{J}), (\textbf{L}), (\textbf{N}) and (\textbf{P})  $dI/dV$ as a function of  $V$ and  $B$  at a fixed value of  $V_{\textrm{bg}}$. The magnetic field scans have been taken at $V_{\textrm{bg}} = -12\, \si{V}$ for device F, at $V_{\textrm{bg}} = -32\, \si{V}$ for device G, at $V_{\textrm{bg}} = -12\, \si{V}$ for device H, at $V_{\textrm{bg}} = -7.5\, \si{V}$ for device I, at $V_{\textrm{bg}} = -19\, \si{V}$ for device J, at $V_{\textrm{bg}} = -15.5\, \si{V}$ for device K, at $V_{\textrm{bg}} = -5\, \si{V}$ for device L and at $V_{\textrm{bg}} = -30\, \si{V}$ for device M. The different shapes and positions of the lobes are due to the slightly different NW and Al shell dimensions. However, neither ZBPs nor other subgap features are observed.}}
\end{figure}

\begin{figure}
\includegraphics[scale=1]{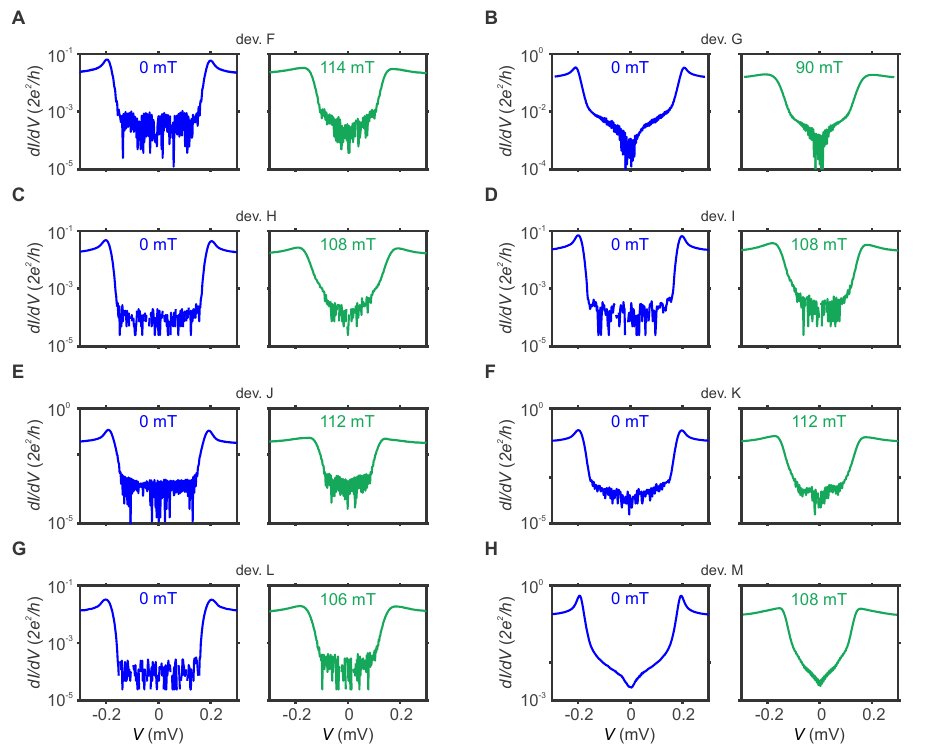}
\caption{\label{fig:figure1_part2_sup} \textbf{Line traces of short junction devices at the center of the zeroth and first lobe.}  \small{ (\textbf{A}) Line-cuts taken from Fig.~\ref{fig:figure_sup1}B. (\textbf{B}) Line-cuts taken from Fig.~\ref{fig:figure_sup1}D. (\textbf{C}) Line-cuts taken from Fig.~\ref{fig:figure_sup1}F. (\textbf{D}) Line-cuts taken from Fig.~\ref{fig:figure_sup1}H. (\textbf{E}) Line-cuts taken from Fig.~\ref{fig:figure_sup1}J. (\textbf{F}) Line-cuts taken from Fig.~\ref{fig:figure_sup1}L. (\textbf{G}) Line-cuts taken from Fig.~\ref{fig:figure_sup1}N. (\textbf{H}) Line-cuts taken from Fig.~\ref{fig:figure_sup1}P.}}
\end{figure}

\begin{figure}
\includegraphics[scale=1]{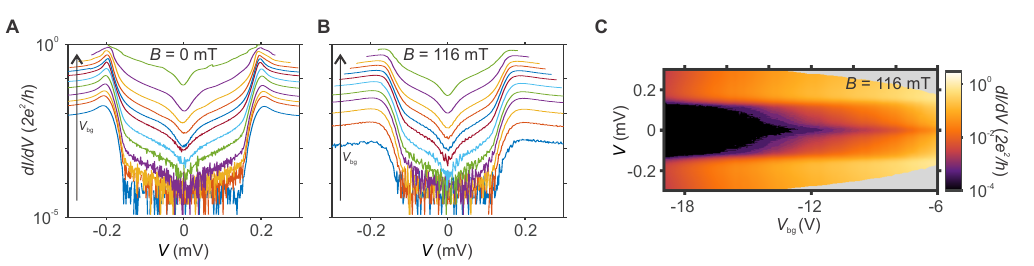}
\caption{\label{fig:figure1_part3_sup} \textbf{Subgap conductance evolution for  device A as a function of the backgate voltage}. \small{(\textbf{A}) Line-cuts extracted from the measurement shown in Fig.~1C, ranging from $V_{\textrm{bg}} = -20.4\, \si{V}$ (lower trace)  to $V_{\textrm{bg}} = -6\, \si{V}$ (upper trace), equally spaced. (\textbf{B}) Line-cuts taken at $B = 116$ mT, i.e. in the middle of the first lobe. The line-cuts range from $V_{\textrm{bg}} = -19.6\, \si{V}$ (lower trace)  to $V_{\textrm{bg}} = -6.4\, \si{V}$ (upper trace), equally spaced. No signature of MZMs is observed independently of the junction transparency. (\textbf{C}) $dI/dV$ as a function of  $V$ and  $V_{\textrm{bg}}$ for device A, at $B = 116\, \si{mT}$, namely in the middle of the first lobe. The line-cuts of (B) are taken from (C).
}}
\end{figure}

%\floatsetup[figure]{capposition=bottom}

\begin{figure} [h]
\includegraphics[scale=1]{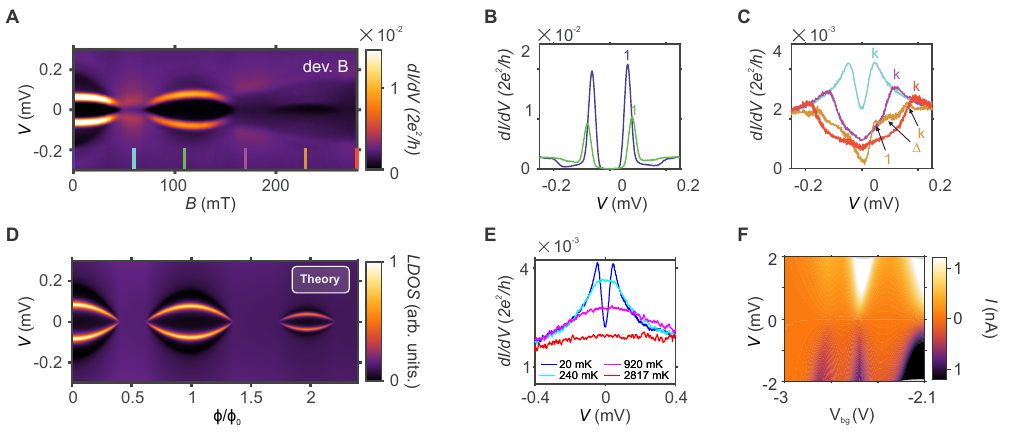}%
\caption{\label{fig:figure_sup7} \textbf{Kondo effect in the Little-Parks regime with a doublet GS from device B.} \small{ (\textbf{A}) Same plot as Fig.~3D but without the analytical solution superimposed. (\textbf{B}) and (\textbf{C}) Line traces extracted from (A) for different values of $B$. The line traces in (B) show that ABSs in the doublet GS, indicated with 1, increase their energy with magnetic field. In the absence of magnetic field, $\zeta \approx \pm 63 \, \si{\micro V}$ (deep blue trace), while at $B = 110 \, \si{mT}$ (green trace)   $\zeta \approx \pm 80 \, \si{\micro V}$, even though the superconducting gap is smaller at finite $B$. (\textbf{C}) Line traces showing the presence of a split Kondo peak (K) in the destructive regimes, namely at $B = 60 \, \si{mT}$ (cyan trace), at $B = 170 \, \si{mT}$ (purple trace) and at $B = 280 \, \si{mT}$ (red trace). Furthermore, the Kondo peaks coexist with superconductivity and ABSs in the 2L (orange trace taken at $B = 230 \, \si{mT}$). (\textbf{D}) Numerical simulation of the $dI/dV$ versus bias voltage $V$ and magnetic flux $\phi/\phi_0$ in a QD-S (modelled as a superconducting Anderson model in the Hartree-Fock approximation), with the SC in the destructive LP regime and the QD in a doublet GS (parameters can be found in Table \ref{params}). (\textbf{E}) $dI/dV$ as a function of $V$ at $B = 60 \, \si{mT}$ for different temperatures. The split Kondo peaks broaden and decrease their conductance forming a ZBP at $T = 240\, \si{mK}$, while they completely disappear for $T = 2817\, \si{mK}$. (\textbf{F}) Current, $I$, as a function of $V$ and $V_{\textrm{bg}}$ for an out-of-plane magnetic field of $50 \, \si{mT}$ that destroys superconductivity revealing Coulomb diamonds. The charging energy is $\approx  2.5 \, \si{meV}$.}}
\end{figure}

%\floatsetup[figure]{capposition=beside,capbesideposition={center,right}}

\begin{figure}[h]
\includegraphics[scale=1]{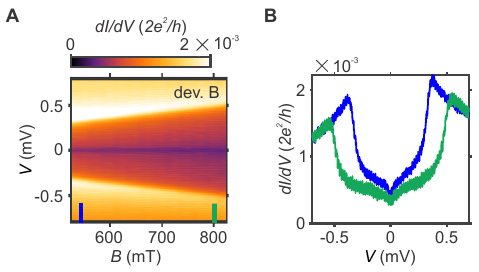}%
\caption{\label{fig:figure_sup5}  \textbf{Kondo data for higher magnetic fields, from device B.} \small{
(\textbf{A}) $dI/dV$ as a function of  $V$ and  $B$ for $V_{\textrm{bg}} = -2.62\, \si{V}$, hence in the doublet GS. (\textbf{B}) Line traces extracted from (A) for $B = 550 \, \si{mT}$ (blue trace) and for  $B = 800 \, \si{mT}$ (green trace).}}
\end{figure}

%\floatsetup[figure]{capposition=bottom}

\begin{figure} [h]
\includegraphics[scale=1]{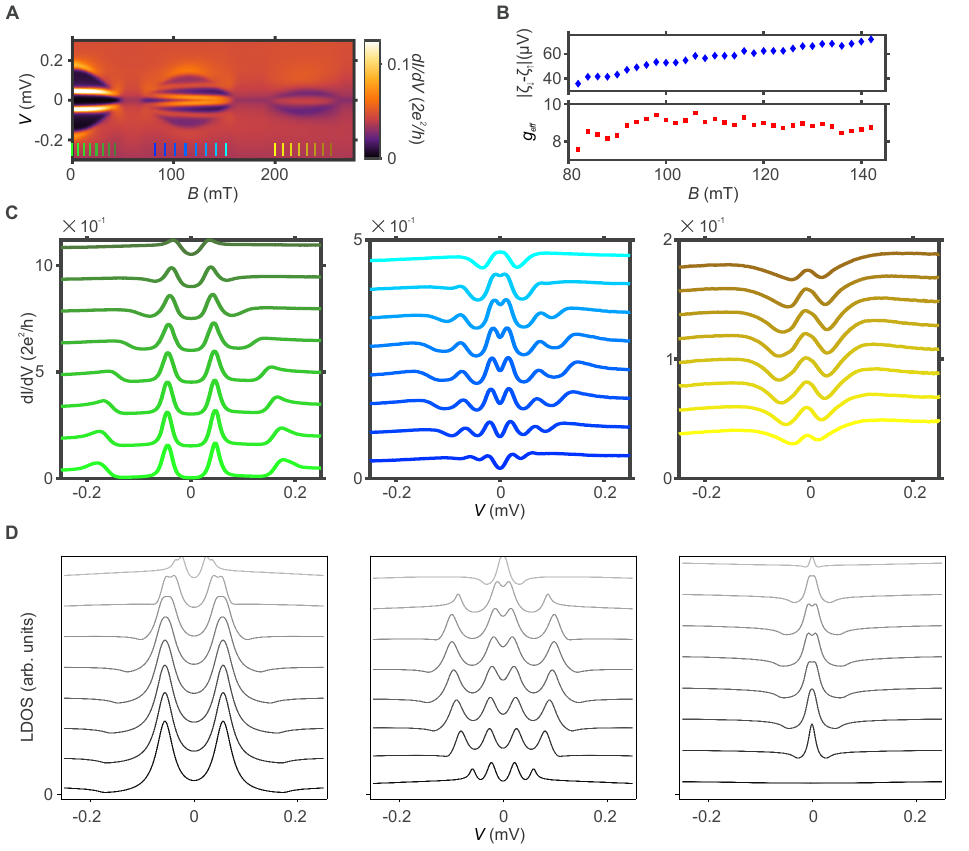}%
\caption{\label{fig:figure_sup9} \textbf{ZBP in the second lobe due to ABSs in device B.}\small{ (\textbf{A}) Same plot as Fig.~4D but with coloured markers. (\textbf{B}) ABSs splitting in the 1L at positive bias versus $B$, (upper panel) and effective g-factor $g_{\rm{eff}}$ extracted from their splitting (lower panel). (\textbf{C}) Line traces extracted from (A) for the 0L, 1L and 2L (from left to right) illustrating how the ABSs merge at zero bias at the end of the 1L.  The lower trace of the 0L is taken at $B =0$ and the upper one is taken at $B = 42 \, \si{mT}$. They are shifted by $0.3 {e^{2}}/{h}$ with respect to each other. For the 1L, the line traces are shifted by 0.12 $e^{2}/{h}$ and they range from  $B = 82 \, \si{mT}$ (lower one) to  $B = 152 \, \si{mT}$ (upper one). Finally, for the 2L, the line traces are shifted by $0.04 {e^{2}}/{h}$ and range from  $B = 200 \, \si{mT}$ (lower one) to  $B = 256 \, \si{mT}$ (upper one). (\textbf{D}) Numerical simulation of (C). The parameters can be found in Table \ref{params}.}}
\end{figure}

%\floatsetup[figure]{capposition=bottom}

\begin{figure} [h]
\includegraphics[scale=1]{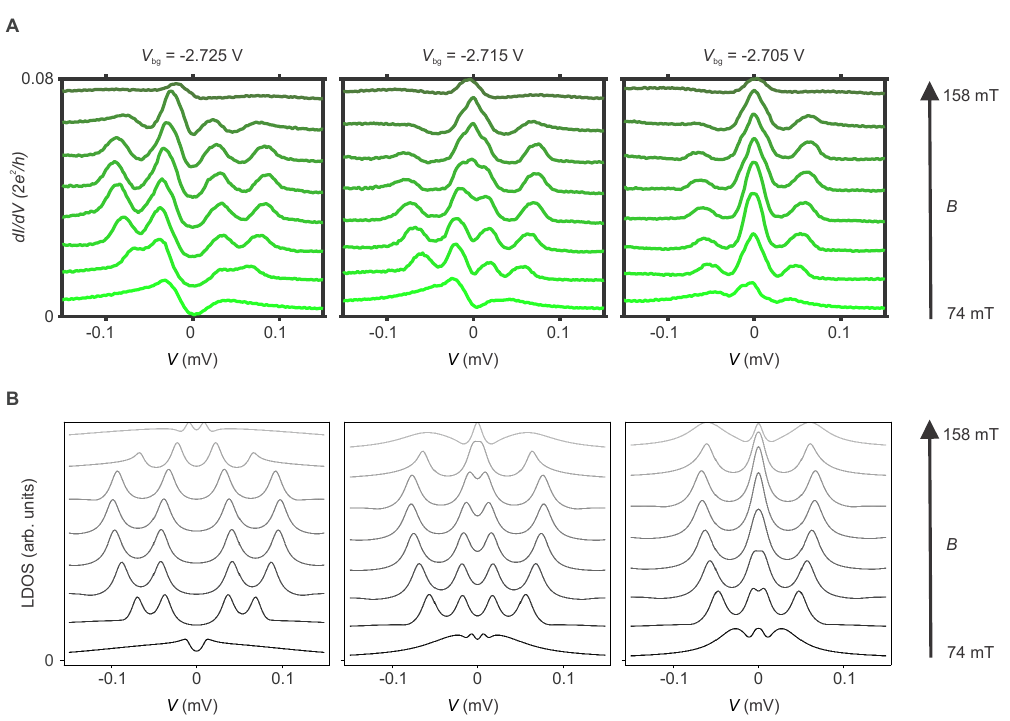}%
\caption{\label{fig:figure_sup8} \textbf{ZBP in the first lobe due to ABSs in device B.}\small{ (\textbf{A}) Line traces extracted from the first lobe of Fig. 4E (left, center and right panels) of the main text}, showing $dI/dV$ as a function of $V$ from $B = 74\, \si{mT}$ (lower one) to $B = 158\, \si{mT}$ (upper one); they are shifted by $0.01 \frac{e^{2}}{h}$ with respect to each other. (\textbf{B}) Numerical simulations of (A) extracted from Fig. 4F (left, center and right panels) of the main text.}
\end{figure}

%\floatsetup[figure]{capposition=beside,capbesideposition={center,right}}

%\floatsetup[figure]{capposition=bottom}

\begin{figure}[h]
\includegraphics[scale=1]{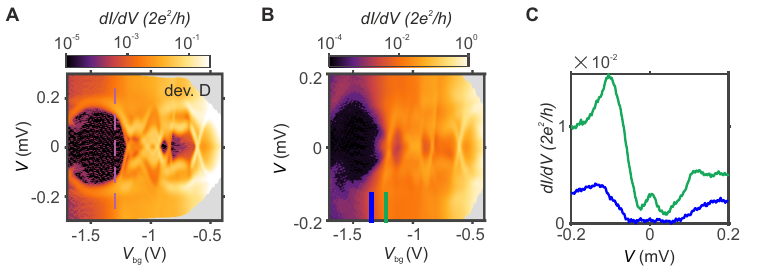}%
\caption{\label{fig:figure_sup3} 
\textbf{Dependence of ZBP on the backgate voltage and additional data for device D.} \small{(\textbf{A}) $dI/dV$ as a function of  $V$ and  $V_{\textrm{bg}}$ at in absence of magnetic field. Multiple ABSs are clearly visible. The dashed pink line indicates where the
magnetic field scan shown in Fig. 5B was taken.
(\textbf{B}) $dI/dV$ as a function of  $V$ and  $V_{\textrm{bg}}$ at $B = 108 \, \si{mT}$, i.e., in the first lobe. The presence of a ZBP depends on  $V_{\textrm{bg}}$, proving that it has no topological origin. (\textbf{C}) Line traces extracted from (B); the blue curve corresponds to $V_{\textrm{bg}} = -1.25 \, \si{V}$ while the green one to  $V_{\textrm{bg}} = -1.34\, \si{V}$.}}
\end{figure}

%\floatsetup[figure]{capposition=beside,capbesideposition={center,right}}
\begin{figure}[h]
\includegraphics[scale=1]{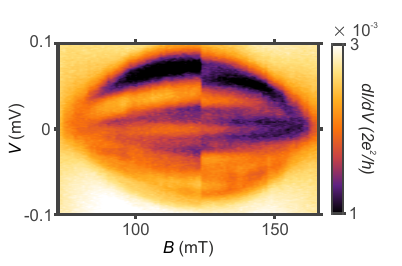}%
\caption{\label{fig:figure_sup10} 
\textbf{Zoom of Fig.~5C.} \small{ $dI/dV$ as a function of $B$ and $V$; same as Fig.~5C but with a finer resolution.}}
\end{figure}

\bibliographystyle{Science}